\documentclass[fleqn,usenatbib,useAMS]{mnras}

\usepackage{newtxtext,newtxmath}

\usepackage[T1]{fontenc}
\usepackage{ae,aecompl}

\usepackage{graphicx}	
\usepackage{amsmath}	
\usepackage{lscape}
\usepackage{ulem}

\newcommand{\msun}{M_{\odot}}
\newcommand{\mstar}{M_*}

\newcommand{\subfind}{{\sc Subfind }}
\newcommand{\sublink}{{\sc SubLink }}

\newcommand{\rperip}{r_{{\rm peri}^-}}
\newcommand{\rperin}{r_{{\rm peri}^+}}

\newcommand{\rapop}{r_{{\rm apo}^-}}
\newcommand{\dtapop}{\Delta t_{{\rm apo}^-}}
\newcommand{\rapon}{r_{{\rm apo}^+}}
\newcommand{\dtapon}{\Delta t_{{\rm apo}^+}}

\newcommand{\dtperip}{\Delta t_{{\rm peri}^-}}
\newcommand{\dtperin}{\Delta t_{{\rm peri}^+}}
\newcommand{\otperi}{t_{{\rm peri}}-t_{{\rm peri}}^{\rm true}}
\newcommand{\orperi}{r_{{\rm peri}}-r_{{\rm peri}}^{\rm true}}
\newcommand{\tmerge}{t_{\rm merge}}
\newcommand{\fmerge}{f_{\rm merge}}
\newcommand{\dtmerge}{\Delta t_{\rm merge}}

\title[Interacting Galaxies in IllustrisTNG - VI] 
{Interacting galaxies in the IllustrisTNG simulations - VI: Reconstructed orbits, close encounters and mergers}

\author[D. R. Patton et al.]{
David R. Patton,$^{1}$\thanks{E-mail: dpatton@trentu.ca}
Lawrence Faria,$^{1,2}$ 
Maan H. Hani,$^{3}$
Paul  Torrey,$^{4}$ 
Sara L. Ellison,$^{3}$
Shivani D. Thakur,$^{1}$ \newauthor and 
Raven I. Westlake$^{1}$ 
\\
$^{1}$Department of Physics and Astronomy, Trent University, 1600 West Bank Drive, Peterborough, ON, K9L 0G2, Canada\\
$^{2}$Department of Physics, Engineering Physics and Astronomy, Queen's University, Kingston, ON, K7L 3N6, Canada\\
$^{3}$Department of Physics and Astronomy, University of Victoria, Finnerty Road, Victoria, BC, V8P 1A1, Canada\\
$^{4}$Department of Astronomy, University of Virginia, 530 McCormick Road, Charlottesville, VA 22903, USA\\
}

\pubyear{2024}

\begin{document}
\label{firstpage}
\pagerange{\pageref{firstpage}--\pageref{lastpage}}
\maketitle

\begin{abstract}
Cosmological simulations have been used to study interacting galaxies
as a function of galaxy pair separation, enabling comparisons with observational studies of galaxy pairs.
The study of interacting galaxies as a function of time (i.e. merger stage) has mostly been
limited to high resolution merger simulations, due to the poor time sampling available in cosmological simulations.
Building on an earlier study of galaxy pairs in the IllustrisTNG cosmological simulations, we reconstruct the orbits of
galaxy pairs involving massive galaxies ($M_* > 10^{10}\msun$) at redshifts of $0 \leq z < 1$, 
using a novel kinematic interpolation scheme to model the orbits in between the IllustrisTNG snapshots (which are separated
by 162 Myr on average).  
We assess the accuracy of these interpolations using a pre-existing suite of merger simulations, and find that
kinematic interpolations provide a remarkable improvement in accuracy compared with interpolations that use
only radial separations or 3D positions.
We find that nearly 90 per cent of the closest pairs ($r < 25$ kpc) have had a pericentre encounter within the past Gyr.
Many of these close pairs are found on rapidly shrinking orbits, and roughly
85 per cent of these pairs will merge within 1 Gyr.
However, approximately 3 per cent of these close pairs appear to be flyby systems that will never merge.
These reconstructed orbits will be used in future studies to investigate
how and when galaxy properties change during close encounters and mergers between galaxies in IllustrisTNG.
 
\end{abstract}




\section{Introduction}\label{secintro}

Early studies of galaxy properties showed that strongly interacting galaxies have morphologies, colours
and star formation rates that are significantly different from those of relatively
isolated galaxies \citep{arp66,toomre72,larson78,kennicutt87}.  Large statistical studies of galaxy pairs
have subsequently shown clear correlations between pair separation and numerous galaxy properties,
suggesting that the degree to which galaxy properties are perturbed may be closely tied  
to the proximity of the encounters.  For example, the star formation rates of galaxies in close pairs
(with typical projected separations less than about 30 kpc) are found to be enhanced by a factor of
roughly 2-3 with respect to relatively isolated
galaxies \citep{ellison08,scudder12,barrera-ballesteros15,stierwalt15,pan18,garduno21,steffen21,shah22}, 
with smaller enhancements extending out to separations of about 150 kpc \citep{patton13,patton20,brown23}.
Galaxies in close pairs have also been found to have lower
gas-phase metallicities \citep{ellison08,rupke10,scudder12,bustamante20,garduno21},
higher asymmetries \citep{patton05,depropris07,casteels14,patton16},
and higher active galactic nuclei (AGN)
fractions \citep{ellison11,ellison13,silverman11,satyapal14,ellison19,steffen23,bickley24} than
matched control samples of galaxies without close companions.
All of these differences have been shown to diminish and then disappear at larger pair separations.

Simulations of merging galaxies have been used to interpret correlations between galaxy properties
and pair separation, in attempts to identify the physical mechanisms by which these perturbed galaxy properties
may arise.
In general, these correlations can be explained by a model in which gravitational and hydrodynamical interactions
trigger morphological disturbances and cause the infall of gas into the central regions of galaxies,
leading to a dilution of gas-phase metallicity,
enhanced star formation, and elevated accretion of gas on the central supermassive black hole
\citep{mihos96,dimatteo05,hopkins08,torrey12,renaud14,moreno19}.
These simulations allow researchers to track changes in galaxy properties as a function of time,
revealing when and how galaxy properties are altered by interactions and mergers.
However, merger simulations are typically carried out on idealized pairs of galaxies on pre-defined orbits
in otherwise isolated
environments.   In reality, the correlations seen in observed samples of galaxy pairs arise within
a wide range of environments \citep{alonso06,ellison10,lin10},
with a diversity of progenitor galaxies, galaxy orientations and orbital configurations.

In recent years, the availability of cosmological simulations has facilitated the theoretical study of more diverse samples
of interacting galaxies, albeit at much lower resolution than
merger simulations \citep{sparre16,bustamante18,blumenthal20,hani20,mcalpine20,patton20,bottrell22,byrne23}.
Cosmological simulations offer an unprecedented opportunity for researchers to bridge the gap between
high resolution simulations of galaxy mergers and large observational samples of galaxy pairs.
In principle, cosmological simulations allow one
to analyze correlations in galaxy properties as a function of pair separation and also 
to analyze changes in galaxy properties {\it as a function of time}, using the same set of simulated galaxies.
In practice, however, the time sampling of cosmological simulations is sufficiently poor (typically $>$ 150 Myr)
that the orbits of interacting galaxies are crudely approximated at best.

The first paper in this series \citep{patton20} reports on the identification of galaxy pairs within the IllustrisTNG cosmological
simulations \citep{nelson19a}.  In that study, the closest companion of each massive galaxy (with stellar masses
of $\mstar > 10^{10} \msun)$ was identified,
with companions required to have at least 10 percent of the stellar mass of the galaxy in question.  
By comparing the specific star formation rates (star formation rates per unit stellar mass; hereafter sSFRs)
of these galaxies with a control sample of relatively
isolated galaxies matched on stellar mass, redshift and environment,
\citet{patton20} found that galaxies with close companions (3D separations $\sim 15$ kpc)
have sSFRs that are enhanced by a factor of about two, with statistically significant enhancements detected out to
separations of 280 kpc.  By repeating their analysis in projected space, \citet{patton20} showed that these results 
are consistent with the correlations between sSFR and projected separation in galaxy pairs from the
Sloan Digital Sky Survey \citep{abazajian09}.  In a subsequent investigation, \citet{brown23} found that
this consistency holds true when dividing the pairs sample based on whether the closest companion
is star forming or passive.
Additional papers in this series have reported on the star formation rates and AGN activity of paired and post-merger galaxies
in IllustrisTNG \citep{hani20,quai21,byrne23,byrne24}.

In this study, we aim to set the stage for the investigation of interacting galaxy properties as a function of time
(rather than as a function of pair separation) by reconstructing the orbits of the IllustrisTNG galaxy pairs
identified by \citet{patton20}.
Our initial goals include the identification and characterization of the significant past and future close encounters
that occur between the galaxies in each pair.  For pairs that subsequently merge, we also determine the time of the merger.
These reconstructed orbits can be used to determine how often close encounters and mergers occur between galaxies
and their closest companions.  In addition, these orbits can be used to determine if 
the differences between paired galaxies and their controls are indeed triggered by close encounters,
as is typically assumed in studies of observed galaxy pairs.  This will provide
a direct link between studies of galaxy pairs in observations and simulations, facilitating the interpretation
of galaxy pairs detected in observational studies.

In the following section, we outline our procedure for reconstructing the orbits of interacting galaxies in IllustrisTNG, including our novel approach for interpolating the orbits in between simulation snapshots.  In Section~3, we use a suite of merger simulations to investigate the accuracy of our interpolations.  In Section~4, we describe the diversity of orbits in our sample, and quantify the time and separation of each galaxy pair's adjacent pericentres and apocentres, along with the time until each pair merges.  In Section~5, we carry out a statistical analysis of the close encounters and mergers that are associated with the closest companion pairs in our sample.  In the final section, we summarize our conclusions and describe future work that will benefit from the methods and/or reconstructed orbits presented in this study.

\section{Reconstruction of Orbits}\label{secreconstruct}

\subsection{The IllustrisTNG Simulations}

The IllustrisTNG project\footnote{http://www.tng-project.org}
consists of a suite of cosmological gravo-magnetohydrodynamical simulations that model dark matter, gas, stars and black holes
within cubic cosmological volumes of approximate dimension 50 Mpc (TNG50), 100 Mpc (TNG100) and 300 Mpc (TNG300), 
adopting a cosmology consistent with the Planck 2015 results \citep{planck16}. 
These simulations produce realistic galaxies in a wide range of environments, and 
they are well-matched to observations \citep{pillepich18a}.
The public release of these simulations is described in \citet{nelson19a} and \citet{nelson19b}.
Additional details of the simulations can be found in \citet{nelson18},
\citet{springel18},
\citet{pillepich18b}, \citet{naiman18} and \citet{marinacci18}.

Our initial goal in this study is to reconstruct the orbits of the closest-companion galaxy pairs identified in the 
TNG100-1 simulation by \citet{patton20}.  The TNG100-1 simulation is the highest resolution run of the IllustrisTNG 100 Mpc volume.
Given our desire to track interactions between large numbers of reasonably well-resolved galaxies spanning a wide range
of environments, this simulation offers the best compromise between resolution and volume of the available IllustrisTNG simulations.

\subsection{The \citet{patton20} TNG100-1 Closest Companion Sample}\label{secpatton2020}

The \citet{patton20} pairs sample was assembled by extending the observational closest companion
methodology of \citet{patton16} to
cosmological simulations.  \citet{patton20} found the closest companion for each massive ($M_* > 10^{10} \msun$) galaxy
at $z < 1$ (snapshots 50-99) in the TNG100-1 simulation,
using the 3D separation (hereafter $r$) between the centre of potential of the two galaxies to make
this determination.  Companions were required to have stellar masses that are at least 10 percent as massive
as the galaxy in question.  With this definition, a companion may be more or less massive than the galaxy in question.

A key limitation in the study of close galaxy pairs in cosmological simulations is the numerical stripping that can occur
when two galaxies have significant overlap in their stellar mass distributions.  In these situations, the IllustrisTNG \subfind
algorithm \citep{springel01,dolag09} has a tendency to assign the outermost stellar mass particles of lower mass companions
to their more massive neighbour \citep{rodriguez15}.
\citet{patton20} demonstrate that this effect can be minimized (although not eliminated) by replacing 
each galaxy's current stellar mass with the maximum stellar mass that it has had within the past 0.5 Gyr.
This revised determination of $M_*$ is used only in cases where the separation of the two galaxies
is less than twice the sum of their stellar half mass radii.  

\begin{figure}
\centerline{\rotatebox{0}{\resizebox{9.0cm}{!}
{\includegraphics{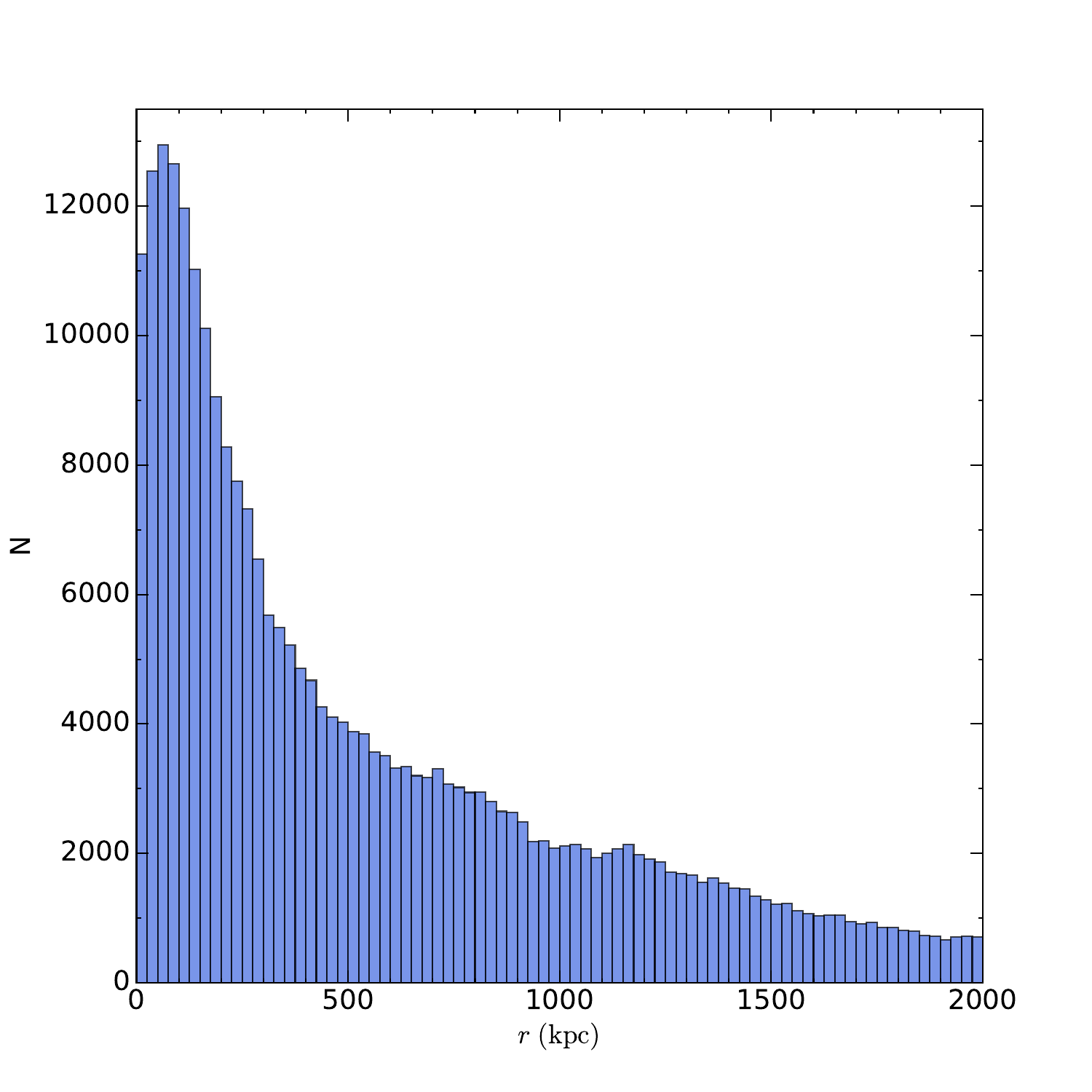}}}}
\caption{A histogram of the 3D distance to the closest companion for massive galaxies ($M_* > 10^{10} \msun$) 
  at $z < 1$ in TNG100-1 is shown, using the closest companion catalog of \citet{patton20}. Each bin is 25 kpc wide.
  Closest companions are required to have at least 10 percent as much stellar mass as the galaxy in question.
  Galaxies that have no such companions within 2 Mpc are not shown.  This sample is likely to be somewhat
  incomplete at the smallest separations (especially $r < 25$ kpc) due to the inability of the \subfind algorithm
  to disentangle highly overlapping galaxy pairs.
\label{figrhist}}
\end{figure}

\citet{patton20} applied these stellar mass and closest companion criteria to all massive galaxies in IllustrisTNG
at $z < 1$,
yielding closest companion information for an average of 5976 massive galaxies at each snapshot. 
In Figure~\ref{figrhist}, we present a histogram of the 3D distance to the closest companion for
massive galaxies ($M_* > 10^{10} \msun$) in TNG100-1, as determined by \citet{patton20}.
This figure includes the 90.2 percent of the sample for which the galaxy's closest companion is closer than 2 Mpc  
(the remainder of the sample is not shown).  The majority of galaxies do not have a companion that would qualify
as ``close'' using the criteria of most studies of interacting galaxies (close pair samples are often restricted to separations
of less than 30 kpc, and few studies of interacting galaxies extend beyond 100 kpc).
In general, the number of galaxies increases with decreasing 3D distance to the closest companion.
However, this trend reverses direction below 50 kpc, with fewer close pairs than would be
expected based on inward extrapolation of the trend at larger separations.
This turnover in the number of pairs may be due to the inability of the \subfind algorithm
to correctly disentangle strongly overlapping galaxy pairs, despite the efforts taken by \citet{patton20} to minimize
the effects of numerical stripping on their sample.

\subsection{Galaxy Velocities}\label{secvel}

We will use the relative velocities of galaxies in IllustrisTNG pairs when reconstructing the orbits of
interacting galaxies.  In order to be useful, these velocities should accurately capture the
relative motion of the centre of each galaxy in a pair at any given snapshot, with each galaxy's centre
being determined by the location of the particle with the lowest gravitational potential energy.

The subhalo velocities that are provided in the IllustrisTNG public data release \citep{nelson19a}
are computed using the mass-weighted velocities of all particles in the subhalo.  These centre-of-mass velocities
are determined mainly by the dark matter in these galaxies.
Unfortunately, the velocity (and position)
of a galaxy's centre of mass can be significantly different from the velocity (and position) of
the centre of the galaxy (i.e. the centre of potential).  This distinction is particularly important
when the galaxy is interacting with or accreting a nearby galaxy \citep{gomez15,vasiliev22,salomon23}. 

We therefore determine our own centre-of-potential velocities for each massive galaxy in TNG100-1,
computing the mass-weighted average velocity in the $x$, $y$ and $z$ directions for all stellar particles
that are located within the galaxy's stellar half mass radius.
We have confirmed that these velocities provide a much better match to the actual motion of each galaxy
than the centre-of-mass velocities available from IllustrisTNG by comparing these velocities with the
change in position of each galaxy over time from one snapshot to the next.  

\subsection{Assembly of Orbit Snapshots}\label{secassembly}

We use the IllustrisTNG \sublink merger trees \citep{rodriguez15}
to trace the orbital history and future of every galaxy pair
in our closest companion sample.  The \sublink merger trees connect \subfind subhaloes
across snapshots, identifying the progenitors and descendants of each galaxy.

We step back in the merger tree for each galaxy by using
its \sublink FirstProgenitor, which is the progenitor with the ``most massive history''
\citep{nelson15}.
A galaxy's FirstProgenitor is usually located at the preceding snapshot.
However, in cases where two galaxies become temporarily blended together at one snapshot,
the \sublink algorithm may intentionally skip a snapshot, resulting in the
FirstProgenitor occurring two snapshots earlier \citep{rodriguez15,nelson15}.
We follow each galaxy's sequence of FirstProgenitors (known as the Main Progenitor Branch)
back in time until we reach a redshift of 1 (SnapNum=50) and/or until the stellar mass of either galaxy
drops below $10^8 \msun$ (below this threshold, there are fewer than 70 stellar particles per galaxy,
  making estimates of stellar mass, position and/or velocity unreliable).

Similarly, we step forward in the merger tree for each galaxy by using
its \sublink Descendant.  Again, while the Descendant is usually
one snapshot forward in time, snapshot skipping will occasionally cause the Descendant
to be two snapshots forward in time.  Where possible, we follow each galaxy's sequence of
Descendants forward in time until a redshift of 0 (SnapNum=99).  

In cases where the galaxies in the pair merge with one another,
the two galaxies will have the same Descendant at all times following the merger.
Following \citet{hani20}, we take the time of the merger (hereafter $\tmerge$)
to be the time of the first snapshot at which
the two galaxies have the same Descendant, while recognizing that the merger may in fact have
occurred at any time between the final pair snapshot and $\tmerge$.
Given that the average time between TNG100-1 snapshots at $0 \leq z < 1$ is 162 Myr \citep{patton20},
this means that the actual merger for any given pair
will presumably have occurred approximately 80 $\pm$ 80 Myr earlier than $\tmerge$.

\begin{figure}
\centerline{\rotatebox{0}{\resizebox{9.0cm}{!}
{\includegraphics{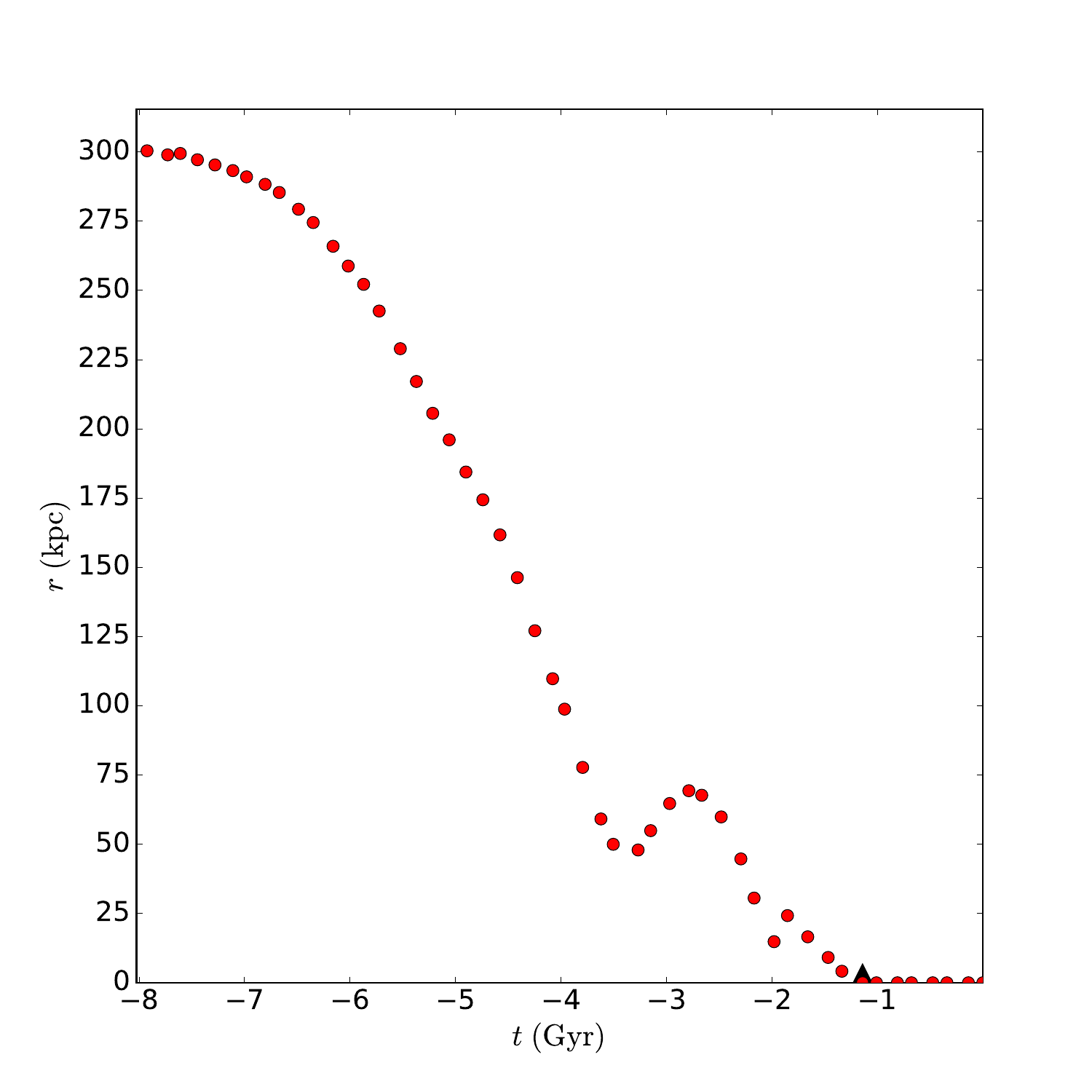}}}}
\caption{3D separation ($r$) is plotted versus time relative to the present day ($t$) for a sample orbit in TNG100-1.
  Red circles denote the separation at each snapshot, and a black triangle indicates the time at which
  this pair merges ($\tmerge$).
\label{figwide}}
\end{figure}

In Figure~\ref{figwide}, we illustrate the results of this approach 
by plotting the three-dimensional separation ($r$) versus time relative to the present day (hereafter $t$)
for a sample orbit in TNG100-1, covering the nearly 8 Gyr timespan corresponding to our redshift range
of $0 \leq z < 1$.
This pair has a separation of about 300 kpc at $z \sim 1$, and merges nearly 7 Gyr (and 42 snapshots) later.
For most of this time, this is not a closest companion system; in fact, the first time that this companion becomes
the closest companion is at a lookback time of 2.0 Gyr, when the pair separation is 15 kpc, and the pair is
840 Myr away from merging.  
There are two places in the orbit where the separation increases, indicating that this pair
undergoes at least two pericentre encounters before it merges.

While the series of snapshots depicted in Figure~\ref{figwide} gives us a rough idea of when these two galaxies
begin interacting and when they merge, the relatively sparse time sampling
of the IllustrisTNG snapshots severely limits our ability to identify and characterize
the encounters between these galaxies.  First of all, while it is clear that at least two encounters occur before the merger,
we cannot say precisely when these encounters occur,
nor do we know how close the galaxies were to one another at the pericentres of these encounters.
In addition, it seems likely that we are missing additional close encounters in the latter stages of this merging sequence, 
as encounters typically occur more frequently as a merging pair approaches coalescence.
These limitations are a common feature in studies of interacting galaxies in cosmological simulations
\citep[e.g.,][]{blumenthal20,sotillo-ramos22,lokas23}.

In the following section, we address these limitations by introducing
and comparing three different methods of interpolating orbits in between simulation snapshots.
Our objective is to improve our ability to identify and characterize close encounters between interacting galaxies
in IllustrisTNG.

\subsection{Interpolation Between Snapshots}\label{secinterp}

The following section describes three interpolation schemes for reconstructing the orbital path
of a given galaxy pair between two consecutive
snapshots.\footnote{We do not attempt to use adjacent snapshots in our interpolations
(e.g. via natural cubic splines), as this is likely to be of limited benefit in modelling rapidly-changing orbits
using sparsely-sampled snapshots with variable time sampling.}
We define the time of the initial and final snapshots to be $t_i$ and
$t_f$ respectively, and we define the corresponding 3D separations to be $r_i$ and $r_f$.

\subsubsection{1D Interpolation Between Snapshots}

The simplest approach to interpolating between two consecutive snapshots
is a linear interpolation in radial separation ($r$).
We refer to this approach as 1D interpolation.  When applied to a plot of pair separation versus time (such as
the orbit depicted in Figure~\ref{figwide}), this is equivalent to simply drawing a straight line between
each adjacent pair of snapshots.

\subsubsection{3D Interpolation Between Snapshots}

We can improve on simple 1D interpolation by taking full advantage of the three dimensional positional information
that is available for each member of the galaxy pair at each snapshot.
The simplest assumption is that the separation between the two galaxies changes linearly with respect
to time in each of the $x$, $y$ and $z$ directions.
We refer to this approach as 3D interpolation.
Our expectation is that 3D interpolation should be an improvement over 1D interpolation, due to the
incorporation of the additional 3D information.
However, we will test this hypothesis in Section~\ref{secvalidation}.

\subsubsection{6D Interpolation Between Snapshots}

Our 1D and 3D interpolation schemes account for the positions of both galaxies in a given pair, but they do not
take the relative velocity of the system into account.  We now introduce a kinematic interpolation scheme, which uses
the separation and relative velocity of the pair as boundary conditions for integrating between two consecutive 
snapshots.  We refer to this as 6D interpolation, since it requires 3D positions and 3D velocities.
The purpose of this interpolation scheme is to find equations for $\Delta x(t)$, $\Delta y(t)$, and $\Delta z(t)$
(and therefore $r(t)$) that satisfy the boundary conditions.
We compute a cubic Hermite spline in each coordinate direction,
finding a unique third-order polynomial\footnote{In principle, if the initial and final accelerations were also known,
  we could find unique solutions using higher order polynomials.
  However, accelerations are not reported for particles or subhaloes in IllustrisTNG.} 
with respect to time $t$.  For completeness, we now outline this procedure in detail for $\Delta x(t)$.

The pair separation in the $x$-direction, $\Delta x(t)$, can be
written as 
\begin{equation}\label{eqn6dx}
\Delta x(t) = A_x + B_x(t-t_i) + C_x(t-t_i)^2 + D_x(t-t_i)^3,
\end{equation}
where $A_x$, $B_x$, $C_x$ and $D_x$ are constants.
We have four constraints on this equation ($\Delta x_i$, $\Delta x_f$, $\Delta \dot{x}_i$ and $\Delta \dot{x}_f$),
which allows us to solve for the four constants, as follows: 
$$A_x = \Delta x_i$$
$$B_x = \Delta \dot{x}_i$$
$$C_x = 3(\Delta x_f-\Delta x_i)(t_f-t_i)^{-2} - (2\Delta \dot{x}_i + \Delta \dot{x}_f)(t_f-t_i)^{-1}$$
$$D_x = 2(\Delta x_i-\Delta x_f)(t_f-t_i)^{-3} + (\Delta \dot{x}_i + \Delta\dot{x}_f) (t_f-t_i)^{-2}$$

Analogous equations for $\Delta y(t)$ and $\Delta z(t)$ have the same form, and are not shown.
We can then compute $r(t)$ by combining $\Delta x(t)$, $\Delta y(t)$ and $\Delta z(t)$ in quadrature.

\begin{figure}
\centerline{\rotatebox{0}{\resizebox{9.0cm}{!}
{\includegraphics{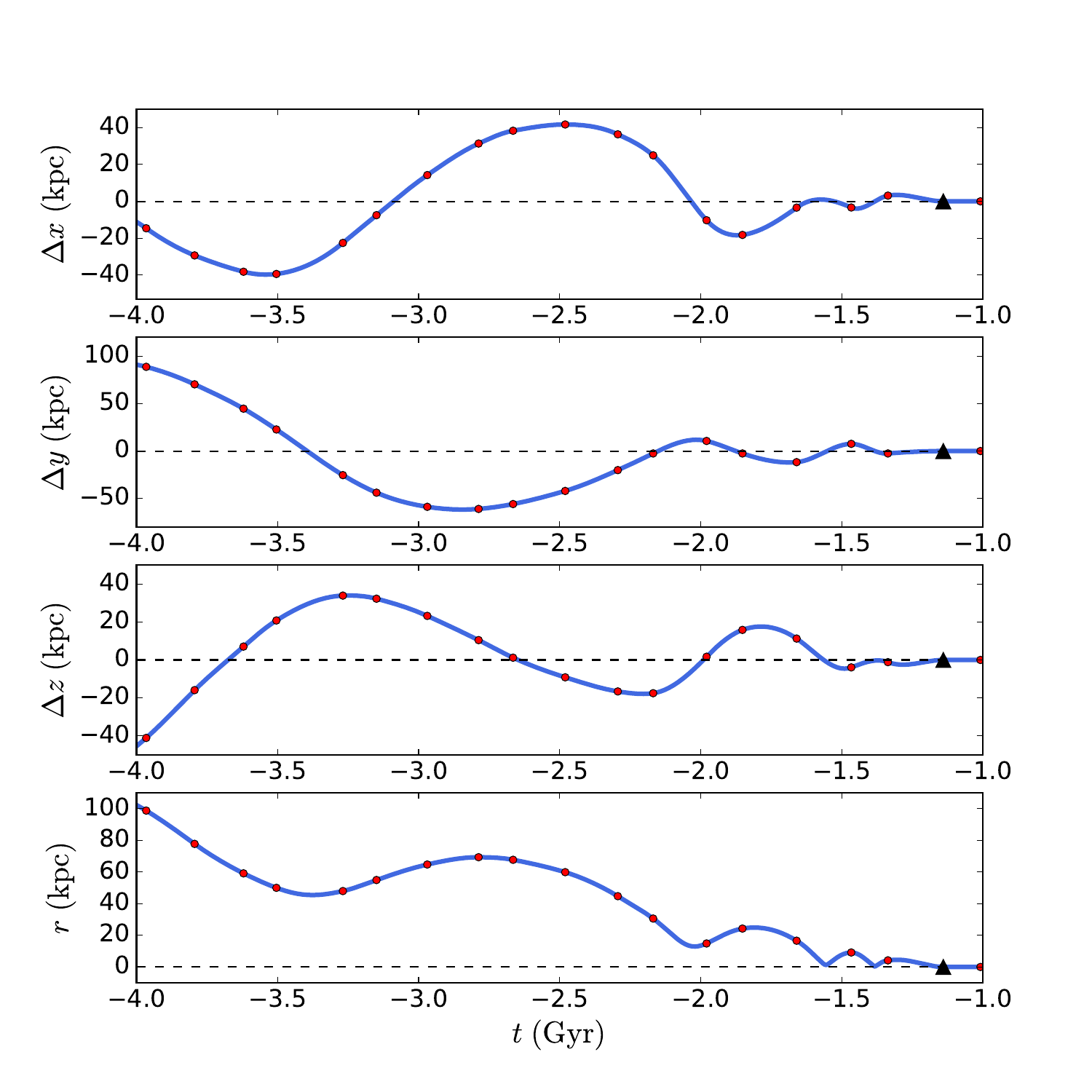}}}}
\caption{Separation is plotted versus time for a sample orbit in TNG100-1.  The upper three panels show
   how the separations along the $x$, $y$ and $z$ axes change with time.  The lowermost panel shows how the
  3D separation changes with time.  Within each panel, red circles denote the separation at each snapshot, 
  the blue solid line depicts the kinematic (6D) interpolation between snapshots, 
  the horizontal dashed line represents a separation that is equal to zero, and the black triangle denotes the merger.  
\label{figrxyz}}
\end{figure}

This kinematic interpolation scheme is illustrated in Figure~\ref{figrxyz}, which shows 
the interpolation in each of $\Delta x$, $\Delta y$ and $\Delta z$, along with the
resulting interpolation in $r$, for the same sample pair shown in Figure~\ref{figwide},
narrowing our focus to the final 3 Gyr of the merging sequence.  This plot shows that
$\Delta x (t)$, $\Delta y (t)$, $\Delta z (t)$ and $r(t)$ all vary smoothly with time,
with the interpolations being continuous at each snapshot (a consequence of
the boundary conditions in both separation and relative velocity).

\subsubsection{1D, 3D and 6D Interpolation of a TNG100-1 Orbit}

\begin{figure}
\centerline{\rotatebox{0}{\resizebox{9.0cm}{!}
{\includegraphics{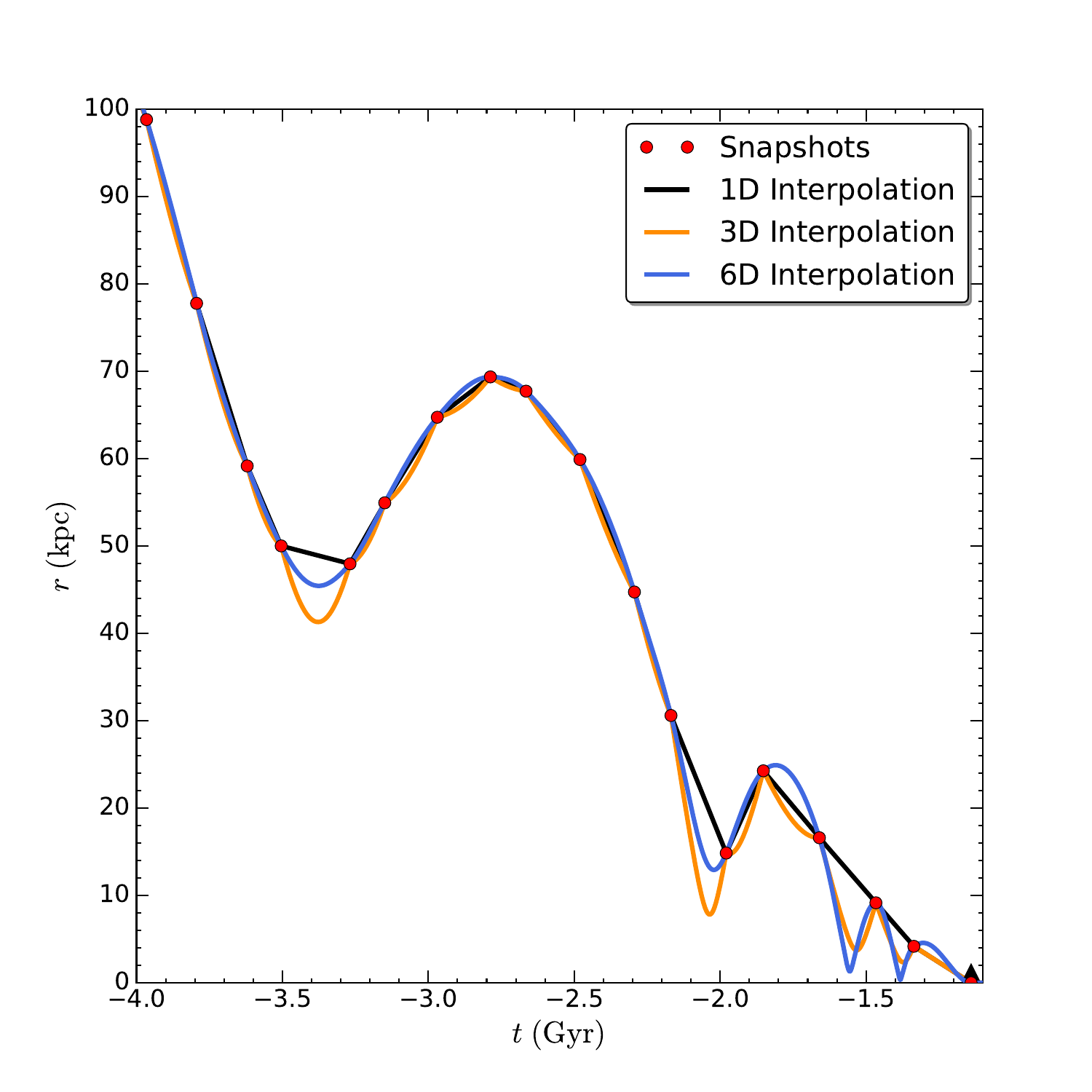}}}}
\caption{3D separation ($r$) is plotted versus time ($t$) for a sample orbit in TNG100-1.
  Red circles denote the separation at each snapshot, and a black triangle indicates the time at which
  this pair merges ($t_{\rm merge}$).
  The black line depicts a linear (1D) interpolation of $r$ with respect to $t$, and yields two close encounters.
  The orange line depicts a 3D interpolation (positional interpolation in each of $x$, $y$ and $z$),
  yielding four close encounters.
  The blue line depicts a 6D interpolation (kinematic interpolation in each of $x$, $y$ and $z$),
  yielding four close encounters.
\label{figclose}}
\end{figure}

We have described three different interpolation schemes (1D, 3D and 6D) for predicting the orbital path between
each pair of snapshots.  In Figure~\ref{figclose}, we apply all three interpolation schemes to the IllustrisTNG orbit
that was depicted in Figure~\ref{figwide}.  In the early stages of this merger sequence, and surrounding the
first apocentre, all three interpolation schemes yield similar results.  On the other hand, these 
interpolation schemes yield strikingly different predictions near each apparent pericentre, as well as during
the late stages of this merging sequence.  

Which of these interpolations is best?
Qualitatively speaking, the 6D interpolations appear to be the most
realistic, with a smooth up-and-down curve that is reminiscent of analogous ``bounce'' plots from idealized merger
simulations \citep[e.g.][]{torrey12,karera22,vasiliev22}.
The smoothness of these interpolations as they pass through each snapshot is a consequence of
our velocity boundary conditions, which force $\dot{x}(t)$, $\dot{y}(t)$ and $\dot{z}(t)$ to be continuous at each snapshot.
This behaviour is what we would expect in the presence of smoothly varying
gravitational forces.  On the other hand, the 3D and 1D interpolations (which do not have velocity boundary conditions)
are continuous but not smooth as they pass through each snapshot.

While there are good reasons to think that our kinematic (6D) interpolations are a major improvement over their 1D or 3D counterparts,
we would like to confirm this by assessing the accuracy of all three interpolation schemes.  More specifically, we would like to quantify
the uncertainties in our predictions of the time and separation of close encounters found using these interpolations,
as this is central to our goal of characterizing the encounters that occur between pairs of galaxies in IllustrisTNG.  
We therefore turn to a pre-existing suite of idealized merger simulations to assess the accuracy of our interpolations,
as described in the following section.

\section{Validation of Kinematic Interpolation Using Merger Simulations}\label{secvalidation}

\subsection{The \citet{patton13} Merger Suite}

To investigate the accuracy of our IllustrisTNG interpolations, we apply our interpolation techniques to a suite of galaxy mergers
that was introduced by \citet{patton13} and further analyzed by \citet{moreno15}.
For the purposes of this validation exercise, the key benefit of these simulations is the superior time sampling
that is available (10 Myr in these merger simulations, versus an average of 162 Myr in IllustrisTNG).

These merger simulations use the
GADGET-2 N-Body/SPH simulation code of \citet{springel05} to model the interaction and merger
of two galaxies (with initial stellar masses of $1.4\times 10^{10}\msun$ and
$5.7\times10^{9}\msun$) placed on a variety of realistic merging orbits.
Five choices of orbital eccentricities (0.85, 0.9, 0.95, 1.0 and 1.05) and impact parameters (2, 4, 8, 12 and 16 kpc) yield
a set of 25 orbits that were chosen to be consistent with the orbital element distribution functions derived from
cosmological simulations by \citet{wetzel11}.  In addition, three sets of merger disc orientations were chosen,
using the $e$, $f$ and $k$ orientations from \citet{robertson06}.  This yields a set of 75 merger simulations that exhibit
a range of initial encounter separations, subsequent apocentre separations, and merger timescales, crudely mimicking the
variety of merging orbits that are found in cosmological simulations.
The output from these simulations was saved every 10 Myr, providing time sampling that is about 16 times better than
for the IllustrisTNG snapshots.  
Please see \citet{patton13} for additional details about this suite of merger simulations.

\subsection{Reconstruction of the Fiducial Orbit}

We begin by examining the merger suite orbit that has 
intermediate values of impact parameter (8 kpc), eccentricity ($0.95$) and orientation ($f$),
calculating the pair separation at every 10 Myr snapshot throughout the interaction and merger sequence.
We take this to represent the true orbit for this fiducial simulation.  This orbit contains four close encounters, with
the merger occurring approximately 2 Gyr after the first encounter.

Next, we create a mock IllustrisTNG orbit for this same fiducial simulation by sampling the orbit
once every 160 Myr (i.e., at every 16th output in the merger simulation).  
We then apply our three interpolation schemes (1D, 3D and 6D) to this set of mock IllustrisTNG snapshots, and  
compare the reconstructed orbit with the true orbit.  We pay particular attention to the accuracy of the interpolations
during the first close encounter, which occurs at $t=0.50$ Gyr and $r =12.3$ kpc in the true orbit, where $t$
is the time relative to the start of the simulation, and $r$ is the 3D separation of the galaxies.

\begin{figure}
\centerline{\rotatebox{0}{\resizebox{9.0cm}{!}
{\includegraphics{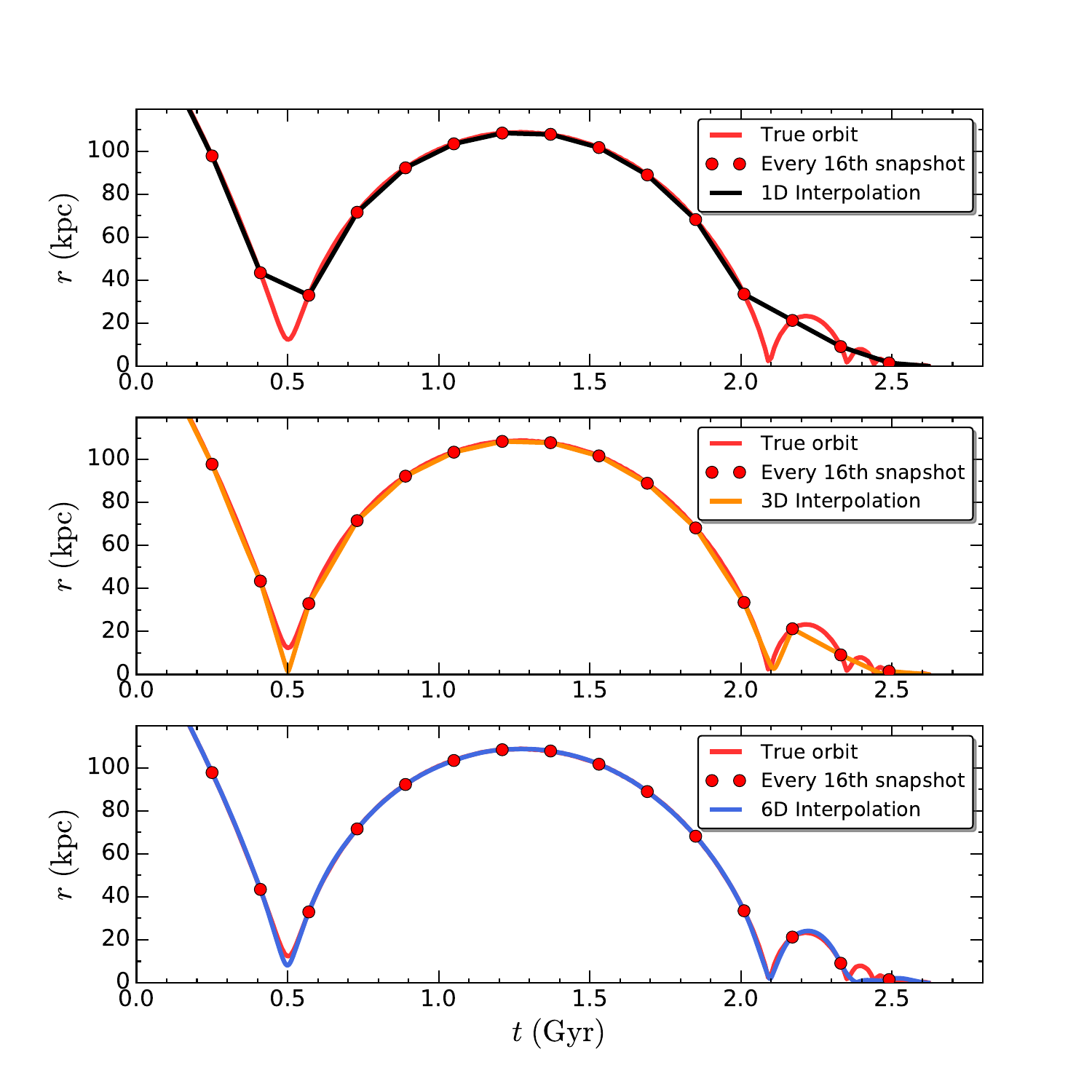}}}}
\caption{Three interpolation schemes are shown for the fiducial orbit from \citet{patton13},
  using a single offset in time.
  In all three panels, the red line shows the true orbit (10 Myr time sampling) and the red circles
  denote every 16th snapshot (with 160 Myr time sampling).  
  The black line in the upper panel show a 1D interpolation between every 16th snapshot.
  The orange line in the middle panel show a 3D interpolation between every 16th snapshot.
  The blue line in the lower panel show a 6D (kinematic) interpolation between every 16th snapshot.
\label{figinterpms}}
\end{figure}

In Figure~\ref{figinterpms}, we plot the 1D, 3D and 6D interpolations for this fiducial orbit, along with the true orbit.
In the upper panel of Figure~\ref{figinterpms}, we see that the 1D interpolation yields a single encounter that corresponds to
the first pericentre of the true orbit.  However, the separation at pericentre is equal to 33 kpc,
which is nearly 3 times larger that the true pericentre separation.  The 1D interpolation also 
does a poor job of identifying the time at which this pericentre occurs, with   
the time of the first pericentre found to be 70 Myr later than the true value.  The 1D interpolation then completely misses 
all subsequent encounters, depicting a monotonic decline in separation until the system merges.

In the middle panel of Figure~\ref{figinterpms}, the 3D interpolation 
is seen to provide a much better fit to the true orbit than the
1D interpolation, capturing the first two encounters reasonably well.
The timing of the first encounter is consistent (within the 10 Myr time sampling) with the true orbit, and the separation
is found to occur at 1.5 kpc (i.e. $\sim 10$ kpc smaller than the true separation).

In the lower panel of Figure~\ref{figinterpms}, the 6D interpolation is seen to provide a substantial improvement over
the 3D interpolation, and a vast improvement over the 1D interpolation.  The 6D interpolation provides an excellent
fit to the first encounter, correctly identifying the time, and estimating the separation to be 8.1 kpc (i.e. 4.2 kpc smaller
than the true separation).  In fact, the 6D interpolation provides an excellent fit to all but the final 0.15 Gyr of the orbit.

\subsection{Generalizing Over All Time Offsets}

For a set of evenly spaced sparsely-sampled snapshots, the
quality of the interpolation at a given close encounter is very sensitive to the proximity of the
snapshots to the encounter.  In particular, if a snapshot happens to fall close to the time of the pericentre,
the resulting interpolation will be much more accurate at (and close to) that pericentre, for all three interpolation techniques.

In Figure~\ref{figinterpms}, we intentionally offset our sparsely-sampled snapshots 
so as to ensure that none of the snapshots fell close to the first or second pericentres,
thereby avoiding the possibility of highly accurate interpolations driven by good luck alone.
We now generalize this approach by reconstructing the fiducial orbit 16 times,
using all 16 possible offsets in time.  This allows us to assess the accuracy
of the interpolations around each pericentre
in an unbiased manner, by including the best case scenarios (when the snapshot falls
very close to the pericentre), the worst
case scenarios (when the pericentre lies midway between the adjacent snapshots), and everything in between.

\begin{figure}
\centerline{\rotatebox{0}{\resizebox{9.0cm}{!}
{\includegraphics{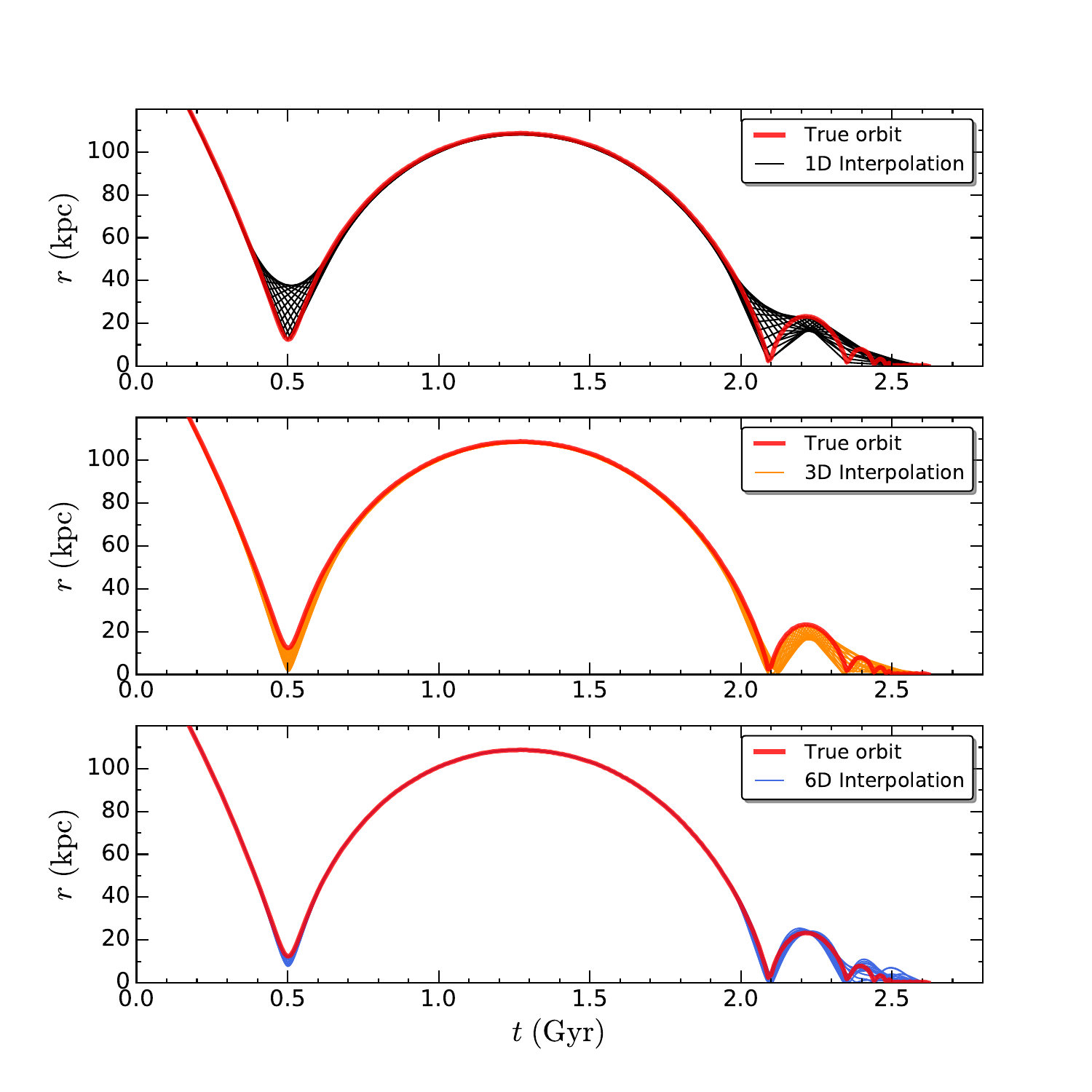}}}}
\caption{Three interpolation schemes are shown for the fiducial orbit from \citet{patton13},
  using all 16 offsets in time.
  In all three panels, the red line shows the true orbit.
  Each black line in the upper panel show a 1D interpolation between every 16th snapshot.
  Each orange line in the middle panel show a 3D interpolation between every 16th snapshot.
  Each blue line in the lower panel show a 6D (kinematic) interpolation between every 16th snapshot.
\label{figinterpmsx16}}
\end{figure}

The results of this comparison are shown in Figure~\ref{figinterpmsx16}.
In the upper panel of this plot, the places where the interpolation provides a poor fit to the true orbit
appear as a black webbed pattern.
The 1D interpolations perform worst in the vicinity of the first and second pericentres,
as well as during the final stages of the merger.  These 1D interpolations {\it overestimate} the separation
at each pericentre, since the encounters identified from 1D interpolations are the local minima found within the available
set of snapshots (i.e. in all but the idealized case of the snapshot landing at the pericentre, the interpolation minimum
will be larger than the true minimum).

In the middle panel of Figure~\ref{figinterpmsx16}, the 3D interpolations are seen to provide better fits to the true orbit 
than the 1D interpolations.  The 3D interpolations almost always {\it underestimate} the pair separation in the vicinity
of the first encounters and in the late stages of the merger sequence.  The 3D interpolations consistently
capture the second encounter, unlike the 1D interpolations.

In the lower panel of Figure~\ref{figinterpmsx16}, the 6D interpolations provide a much better fit to the true orbit
than even the 3D interpolations.  As with Figure~\ref{figinterpms}, we again see that these interpolations usually underestimate
the separation at each pericentre.  Remarkably, most of these 6D interpolations also provide a good fit to the third close encounter.

\subsection{Interpolation Accuracy at the First Pericentre}

One of the primary goals of this study is to identify and characterize close encounters within the orbits of IllustrisTNG galaxy pairs.
While Figure~\ref{figinterpmsx16} provides useful insights in the accuracy of our interpolations for the fiducial orbit from 
\citet{patton13}, we now extend our comparison to the full suite of 75 orbits from this merger suite, focusing on the accuracy of
the interpolations at the first pericentre.  In order to increase the precision of this analysis, we improve the time sampling of
the true orbits by applying our 6D interpolation technique to the merger simulation snapshots (which are separated by 10 Myr).
This allows us to reliably estimate the time of the true pericentre to the nearest Myr (interpolations are much more accurate
within a small time interval).  
For each of the 75 orbits, we then create mock IllustrisTNG orbits (i.e., with 160 Myr time sampling) for all 16 time samplings. 
After reconstructing each of these 1200 mock orbits using 1D, 3D and 6D interpolations, 
we determine the time and separation of the first pericentre (i.e. the first local minimum in the interpolation),
and then calculate the offset of these values from the true time and separation of the first pericentre.  

\begin{figure}
\centerline{\rotatebox{0}{\resizebox{9.0cm}{!}
{\includegraphics{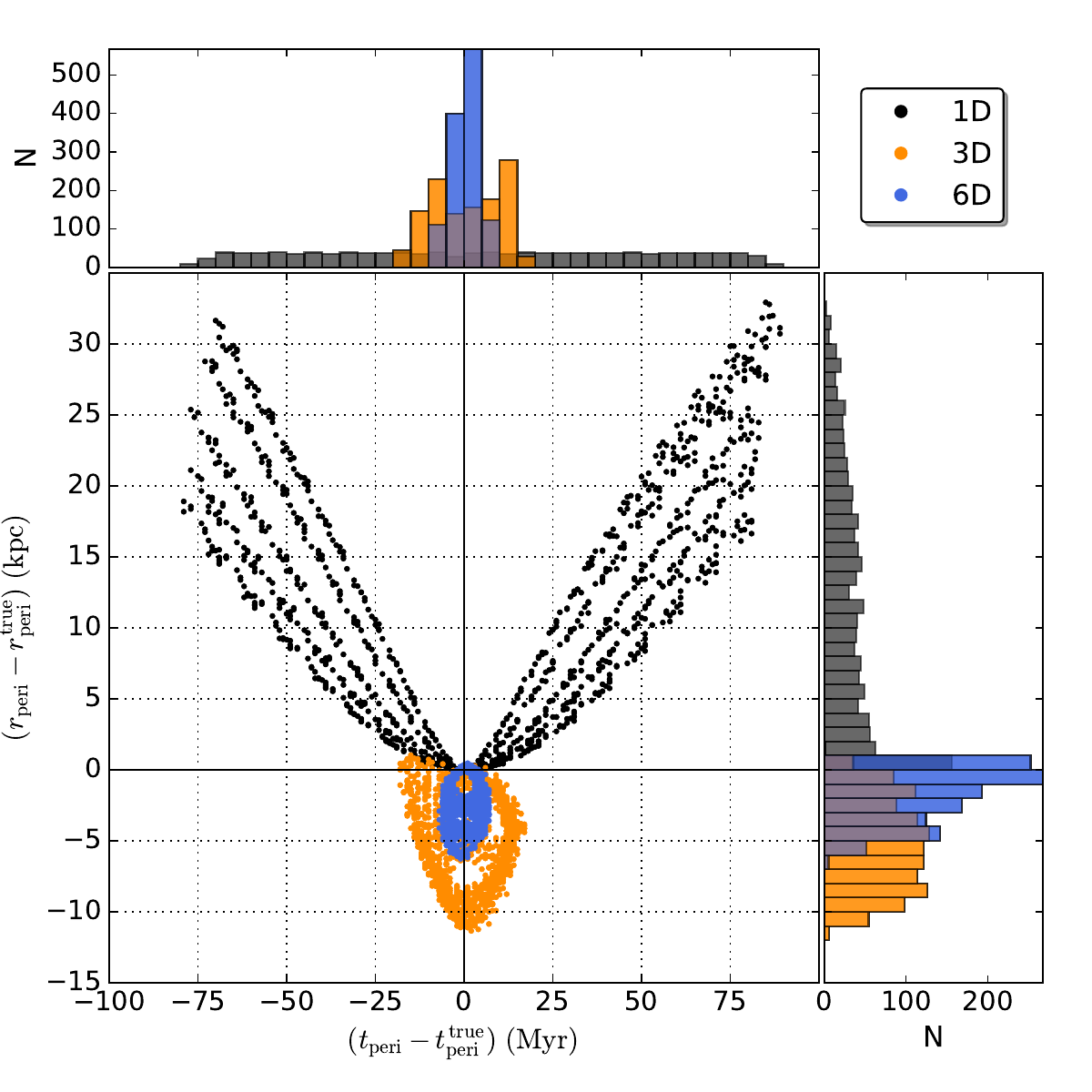}}}}
\caption{The accuracy of the 1D, 3D and 6D interpolation schemes are compared by plotting the
offsets in time and separation at the first pericentre for all 16 time samplings of all 75 orbits from \citet{patton13}.
A perfect interpolation would be located at the origin of this scatter plot.
Colour coding is used to distinguish between 1D interpolations (black), 3D interpolations (orange) and
6D interpolations (blue).  A histogram of $\otperi$ is shown above the scatter plot, and a histogram of
$\orperi$ is shown to the right of the scatter plot.
\label{figperioffhist}}
\end{figure}

In Figure~\ref{figperioffhist}, we plot the offset in pericentre separation ($\orperi$) vs. the offset in pericentre time
($\otperi$) for all 1200 reconstructed orbits.  Above and to the right of this scatter plot, we include histograms of
time and separation offsets respectively.  
In addition, we quantify the time and separation offsets for all three interpolation techniques
by computing the median offsets for each,
along with the standard deviation, reporting these results in Table 1.  
1D interpolations exhibit a wide spread
of offsets in time and separation\footnote{The five distinct sequences seen within the distribution of 1D offsets
  correspond to the five discrete impact
  parameters selected by \citet{patton13}, with the smallest impact parameter (2 kpc) producing the sequence
  with the largest offsets.},
with pericentre separations always overestimated (as expected).
Compared with 1D interpolations, 3D interpolations have much smaller offsets, yielding roughly a factor of five improvement
in time and a factor of three improvement in separation, with separation offsets that are almost always underestimated.
6D interpolations are exceptionally good at identifying the time of the pericentre, with no significant median offset, and
a spread of only 3.3 Myr.  This is $\sim 15$ times better than the 1D interpolations and $\sim 3$ times better than
the 3D interpolations.  6D interpolations also excel at identifying the pericentre separation, with a small (negative) median offset
of 1.4 kpc and a spread of only 1.8 kpc.  

\begin{table}
\centering
\caption{The median offset in time and separation at the first pericentre, computed using 16 time samplings of the
  75 orbits of \citet{patton13}.  The quoted uncertainties are standard deviations.
\label{tabsim}}
\begin{tabular}{cccccc}
\hline
Interpolation&$\otperi$&$\orperi$\\
Method &(Myr)&(kpc)\\
\hline
1D&$5.0 \pm 46.3$&$10.4 \pm 8.7$\\
3D&$1.0 \pm 9.3$&$-5.4 \pm 3.1$\\
6D&$0.0 \pm 3.3$&$-1.4 \pm 1.8$\\
\hline
\end{tabular}
\end{table}

\subsection{Success at Detecting Pericentres}

Broadly speaking, Figure~\ref{figinterpmsx16} indicates that all three interpolation schemes become increasingly inaccurate
during the final stages of the merger, eventually leading to the failure to detect successive pericentres and apocentres,
even for our 6D interpolations.
These trends are consistent with what we would expect when we compare the timescales of the orbit to the time sampling
of the snapshots.  The orbital period (hereafter $\tau$) at any given stage in the merger sequence can be estimated
by computing the time between successive pericentres or apocentres.   In the final stages of the merger sequence, $\tau$ becomes
smaller than the average time between snapshots ($\sim 160$ Myr), posing serious challenges for any interpolation scheme.

In order to identify where our 6D interpolations are no longer able to detect every pericentre,
we carried out a detailed inspection of our
1200 mock IllustrisTNG orbits (with time sampling of 160 Myr), comparing our reconstructed orbits with the true
orbits.  In cases where no more than one pericentre lies between an adjacent pair of snapshots,
we find that the reconstructed orbits are always successful at detecting the pericentre.
On the other hand, if there are two pericentres in between an adjacent pair of snapshots,
the 6D interpolations often fail to detect both pericentres, with the likelihood of success depending on
where the snapshots happen to lie in relation to the pericentres.
These findings indicate that 6D interpolations are reliable at detecting pericentres as long as
the orbital period is greater than the time between snapshots.

Of course, the challenge with interpreting orbits that have been reconstructed from sparely sampled snapshots is that
we cannot accurately determine the orbital period during the late stages of the merger sequence.  In particular,
if the interpolation misses a close encounter, the orbital period may appear to be significantly larger than it is.
To address this issue, we exploit the well-known relationship between orbital period and semi-major axis
(hereafter $a$) that is enshrined in Kepler's third law.  We investigate this relationship in the late stages of the
merger sequence by determining $\tau$ and $a$ following the second, third and fourth pericentres
in the \citet{patton13} merger suite.
In each case, we compute the orbital period by doubling the time between the given pericentre and the subsequent apocentre,
and we find the length of the semi-major axis by taking the average of the pair separation at the given pericentre and the subsequent
apocentre.

\begin{figure}
\centerline{\rotatebox{0}{\resizebox{9.0cm}{!}
{\includegraphics{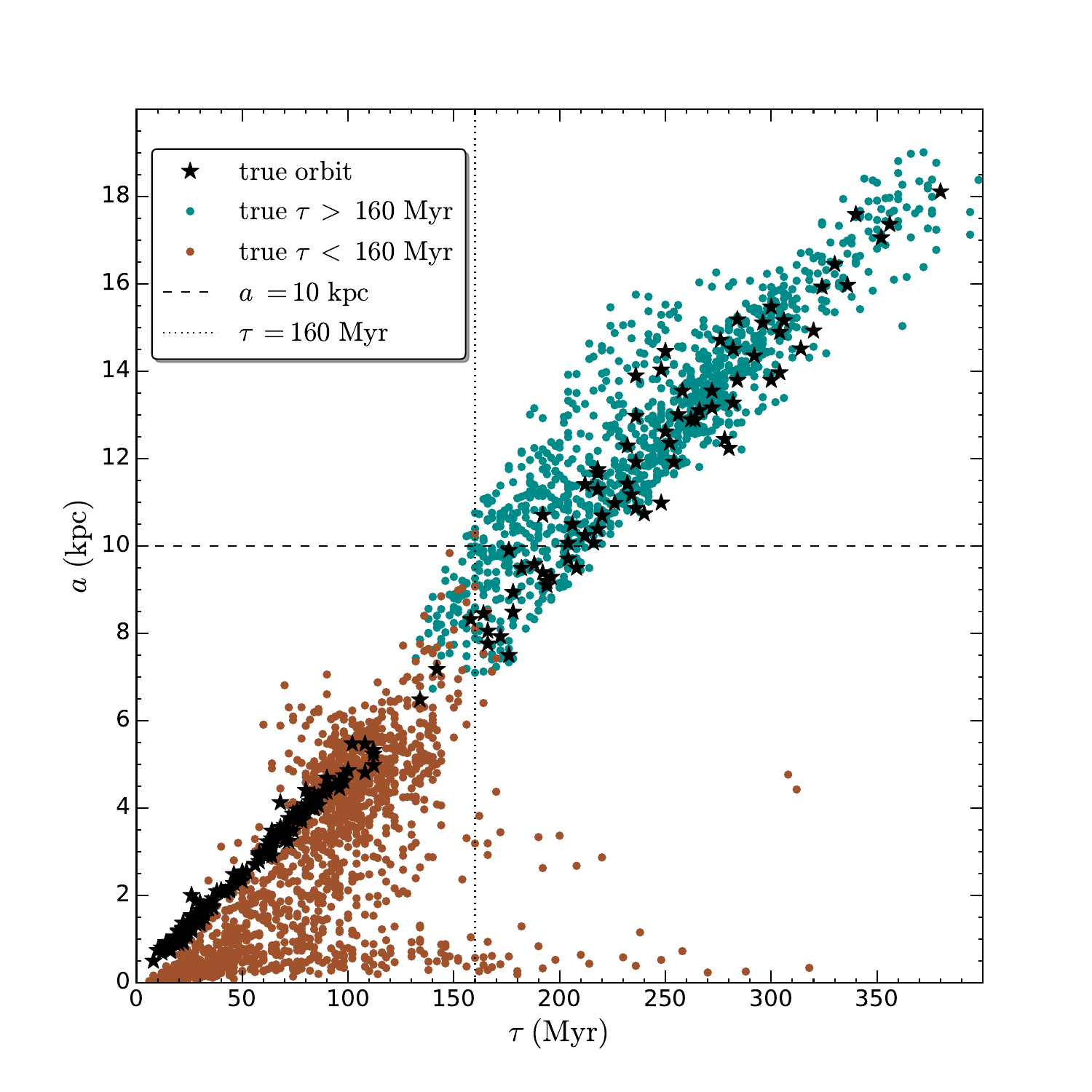}}}}
\caption{Semi-major axis ($a$) is plotted versus orbital period ($\tau$) following the second, third and fourth encounters
  in the 75 orbits from the \citet{patton13} merger suite.  Black stars depict the ``true'' values of $a$ and $\tau$,
  determined using 6D interpolations between snapshots separated by 10 Myr.
  Coloured circles show the corresponding values of $a$ and $\tau$ computed using 6D interpolations
  between snapshots separated by 160 Myr, with green circles for orbits with true $\tau > 160$ Myr
  and brown circles for orbits with true $\tau < 160$ Myr.
\label{figk3}}
\end{figure}

We first apply this methodology to the true orbits in the merger suite (6D interpolations between
10 Myr snapshots), plotting the results as black stars in Figure~\ref{figk3}.
A clear correlation\footnote{This correlation is roughly linear at small separations,
  whereas Kepler's third law has $\tau^2 \propto a^3$.
  This apparent discrepancy is likely due to the extended and overlapping mass distributions of these pairs.}
between
$\tau$ and $a$ is seen, with orbital periods of $\sim$ 10-400 Myr and semi-major axes of $\sim$ 0.5-20 kpc. 
We note that the orbits that have $\tau > 160$ Myr all have $a > $ 7 kpc.   This suggests that if we were able to
restrict our analysis to orbit segments with $a > $ 7 kpc, we could avoid orbital periods of less than 160 Myr.

We next apply this methodology to our 1200 mock IllustrisTNG orbits, which sample the \citet{patton13} merger suite 
using 160 Myr snapshots.  We plot the corresponding values of $\tau$ and $a$ in Figure~\ref{figk3},
colour-coding the data based on whether the true orbital period is greater than 160 Myr (green circles)
or less than 160 Myr (brown circles).  The sparse time sampling of the reconstructed orbits
induces a substantial spread in the estimates of both $\tau$ and $a$, particularly in the lower part of the plot,
where the estimated orbital periods are much larger than their true orbital periods.
However, in all cases where the estimated semi-major axis is greater than 10 kpc,
the true (and estimated) orbital period is greater than 160 Myr.  This suggests the following
rule of thumb: {\it 6D interpolations based on snapshots separated by 160 Myr should reliably
detect all pericentres within snapshot intervals where the average pair separation is at least 10 kpc}.

This rule of thumb confirms our impression from Figure~\ref{figinterpmsx16} that 6D interpolations typically fail
only in the very final stages of the merger sequence, when the average separation indeed
becomes smaller than $\sim$ 10 kpc.
In practice, the quality of the interpolations will depend on the details of a given orbit, including the dynamical masses
of the galaxies (which affect the slope of the $a$-$\tau$ correlation) and the specific spacing of the
IllustrisTNG snapshots (which varies from 123 to 236 Myr at $z < 1$).  We note also that this rule of thumb
scales roughly linearly with the available time sampling
(e.g., with 80 Myr snapshots, we have confirmed that pericentres can be reliably detected if $a > 5$ kpc).
 
In summary, our 6D interpolation technique provides a dramatic improvement in accuracy over 3D and especially 1D
interpolation between snapshots, and is successful at detecting encounters as long as the average pair separation is
$\gtrsim$ 10 kpc.  It therefore seems likely that 6D interpolations between IllustrisTNG snapshots should be
capable of identifying most of the meaningful encounters between galaxies, while yielding estimates of the
time and separation of these pericentres that are much more accurate than for 1D or even 3D interpolations.

\section{Description and Characterization of Reconstructed Orbits}

We have shown that our 6D interpolation scheme is successful at identifying many close encounters
in merger simulations with time sampling of 160 Myr.  We now apply this technique to
the orbits of galaxy pairs in IllustrisTNG, which have an average time between snapshots
of 162 Myr at $z < 1$.  We apply this technique to all TNG100-1 closest companion galaxy pairs with $r < 2$ Mpc,
creating reconstructed orbits in every case where the galaxy pair (or corresponding post-merger)
is identified at two or more snapshots.\footnote{In cases where one of the galaxies drops below $10^8 \msun$
or a merger occurs immediately after $z=1$ (SnapNum=51), the pair may be seen at a single snapshot,
yielding minimal information for that orbit.}
This procedure yields a sample of 32,319 unique reconstructed orbits.

\subsection{Multiple Views of a Sample Reconstructed Orbit in IllustrisTNG}

Each reconstructed orbit provides a record of the 3D separation ($r$) of a galaxy pair at each available IllustrisTNG snapshot,
as well as the interpolated separation at each Myr in between snapshots.  In addition to $r(t)$, the interpolations provide estimates
of the separation in each of the $x$, $y$ and $z$ directions (see Equation~\ref{eqn6dx}).
Finally, by differentiating Equation~\ref{eqn6dx} with respect to time, we can determine the velocity in each of the
$x$, $y$ and $z$ directions, and we can combine these quantities in quadrature to find the relative velocity of each pair
at every Myr along the reconstructed orbit.  Taken together, these quantities provide a wealth of information about
the shape of the orbit and of the relative motion of each galaxy pair.  

\begin{figure}
\centerline{\rotatebox{0}{\resizebox{9.0cm}{!}
{\includegraphics{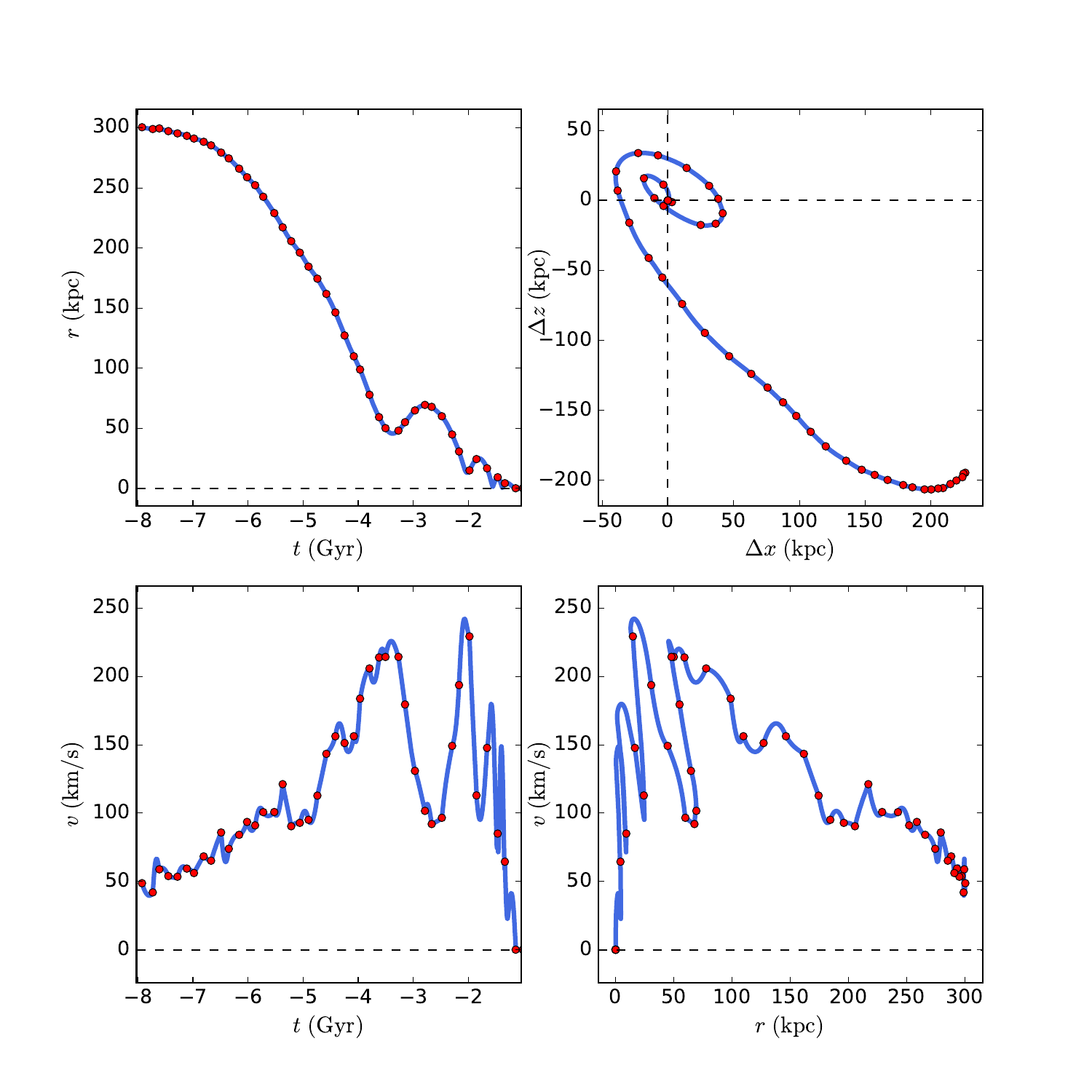}}}}
\caption{A sample TNG orbit is depicted.
  3D separation is plotted versus time in the upper left panel.
  3D relative velocity is plotted versus time in the lower left panel.
  Relative position within the $x-z$ plane is plotted in the upper right panel.
  3D relative velocity is plotted vs. 3D separation in the lower right panel.
  Within each panel, red circles denote the orbital properties at each snapshot, while the blue solid line depicts the
  kinematic interpoloation between snapshots.  In all panels, horizontal and vertical dashed lines denote
  separations or velocities that are equal to zero.
\label{figrvxyzproto}}
\end{figure}

In Figure~\ref{figrvxyzproto}, we provide a visualization of this information by plotting multiple views of the
TNG100-1 orbit that was initially portrayed in Figure~\ref{figwide}.
In the upper left hand panel,
3D separation is plotted versus time, extending the 6D interpolation from Figure~\ref{figclose} 
to the full time interval shown in Figure~\ref{figwide}. In the upper right panel, we plot $\Delta z$ versus
$\Delta x$, showing the trajectory of this orbit within the $x-z$ plane.  The orbital path that is seen is approximately
elliptical in shape, with the size of the ellipse shrinking during each close encounter.
In the lower left panel, 3D velocity is plotted versus time, showing that the velocity peaks at around the time
of each close encounter.  In the lower right panel, 3D velocity is plotted vs. 3D separation,
showing the interplay between these two quantities throughout the merging sequence.

\subsection{Identification of Pericentres and Apocentres}\label{secperiapo}

One of the goals of this study is to identify and characterize the close encounters that are
associated with each galaxy pair.  In principle, it is straightforward to extract this information from our
reconstructed orbits, with each local minimum (maximum) corresponding to a pericentre (apocentre).
In practice, however, some orbits are much more complex than the one depicted in Figure~\ref{figrvxyzproto},
with small variations in separation producing subtle local minima and local maxima that are unlikely to
be associated with any meaningful changes in galaxy properties.  We therefore choose to restrict our analysis
by requiring each pericentre to have a 3D separation that is at least 10 percent smaller than the
3D separation at the adjacent (preceding and following) apocentres.  The locations of the corresponding pericentres
and apocentres are identified on the reconstructed orbits that follow later in this manuscript.

Having identified pericentres and apocentres within each reconstructed orbit, we then 
situate each closest companion pair with respect to adjacent encounters within its orbit.
Of paramount importance is the identification of the most recent pericentre encounter between the two galaxies, as
that encounter may have had a strong influence on the properties of the galaxies in the pair.
Using the reconstructed orbit, we compute the time since the most recent pericentre (hereafter $\dtperip$)
and the separation of the pair at that pericentre (hereafter $\rperip$).

Given the possibility that a given pair may be more strongly influenced by an imminent close encounter than
any past encounters, we also compute the time until the next pericentre (hereafter $\dtperin$), along with the separation
at that pericentre (hereafter $\rperin$).  Finally, we compute the corresponding times and separations
of the adjacent apocentres (hereafter $\dtapop$, $\rapop$, $\dtapon$ and $\rapon$).

\subsection{The Diversity of Reconstructed Orbits in TNG100-1}\label{reconstruct}

The TNG100-1 orbit portrayed in Figure~\ref{figrvxyzproto}, like many other reconstructed orbits in our sample,
is reminiscent of those seen in high resolution simulations of idealized galaxy mergers, whereby two galaxies
merge in an environment that is devoid of other galaxies or large scale structure.  In these scenarios,
the galaxies in a pair undergo a series of progressively closer encounters, typically ending with a merger
within 1-2 Gyr \citep{dimatteo07,cox08,lotz08,hopkins13,moreno21}.
In reality, however, interacting galaxies do not normally evolve in isolation, and neighbouring galaxies or structures 
may exert a strong dynamical influence on a given galaxy pair \citep{moreno13,an19,contreras-santos23}.
These more complex interactions between galaxies arise naturally in IllustrisTNG and other cosmological simulations,
and can be seen in many of the reconstructed orbits in our sample.

\begin{figure}
\centerline{\rotatebox{0}{\resizebox{10.0cm}{!}
{\includegraphics{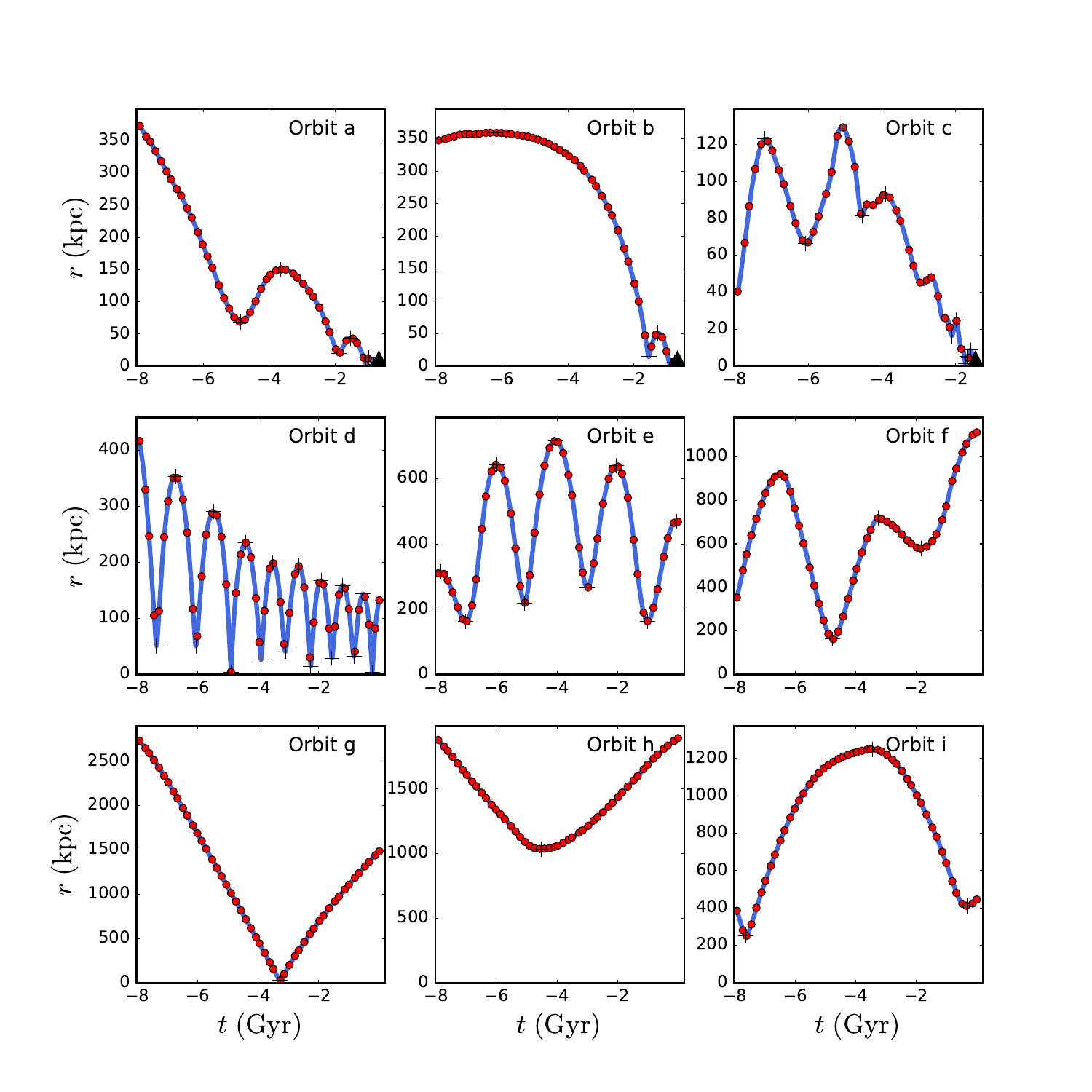}}}}
\caption{A sampling of orbits illustrates the diversity of interactions that are present in our sample of reconstructed orbits.
  In each panel, filled red circles denote snapshots, the solid blue line shows the 6D interpolation,
  {\bf +} symbols locate pericentres and apocentres, and a black triangle
  denotes the time of the merger (if one occurs).
  The top row (Orbits a-c) depicts pairs that merge, with a variety of pre-merger orbits.
  The middle row includes pairs that are unlikely to merge soon (if ever) despite having experienced multiple encounters (Orbits d-f).
  The bottom row shows pairs with flyby or widely spaced encounters (Orbits g-i).
\label{figdiversity}}
\end{figure}

In Figure~\ref{figdiversity}, we illustrate the diversity of closest-companion orbits in IllustrisTNG
by plotting 3D separation versus time for a representative set of 9 reconstructed orbits.  
The top row consists of three pairs that merge,
including a fairly typical merging pair (Orbit a),
a pair that merges rapidly after its initial infall (Orbit b),
and a pair with a complex pre-merger sequence (Orbit c).
The middle row depicts three pairs that have experienced 
multiple encounters but which appear unlikely to merge soon (if ever).  These include
a long-lived pair whose orbit is decaying slowly (Orbit d),
a long-lived pair whose orbit does not appear to be decaying (Orbit e), 
and a pair that is undergoing strong interactions with other galaxies in its vicinity (Orbit f).
The bottom row of Figure~\ref{figdiversity} shows a very close flyby encounter (Orbit g), a much wider flyby encounter (Orbit h),
and an orbit that has a pair of encounters that are separated by more than 7 Gyr (Orbit i).

\subsection{Single Companions Versus Multiple Companions}

Throughout this analysis, we have described the orbits of closest-companion galaxy pairs
which -- by definition -- contain
exactly two galaxies at all times.  While some of the orbits in Figure~\ref{figdiversity} are described as being influenced
by additional neighbouring galaxies, this may nevertheless have given the reader the mistaken impression that
we are modelling the orbits of isolated galaxy pairs.
In reality, galaxies often have many companions at a range of separations.
For example, in Figure 4 of \citet{patton20}, the average number of companions
within 2 Mpc ($N_2$) ranges from 7 to 22 for galaxies in our TNG100-1 sample that have a closest companion
within 1 Mpc.
In addition, the ``closest companion'' designation may switch from one galaxy to another at different snapshots.
This can occur if there are two or more companions with orbits whose radial extents overlap with one another, if the
change in mass of a neighbouring galaxy moves it above or below the 10 per cent threshold used to identify companions
(see Section~\ref{secpatton2020}), or if the closest companion merges with the galaxy in question, thereby
passing on the title of closest companion to the next closest companion.

\begin{figure}
\centerline{\rotatebox{0}{\resizebox{9.0cm}{!}
{\includegraphics{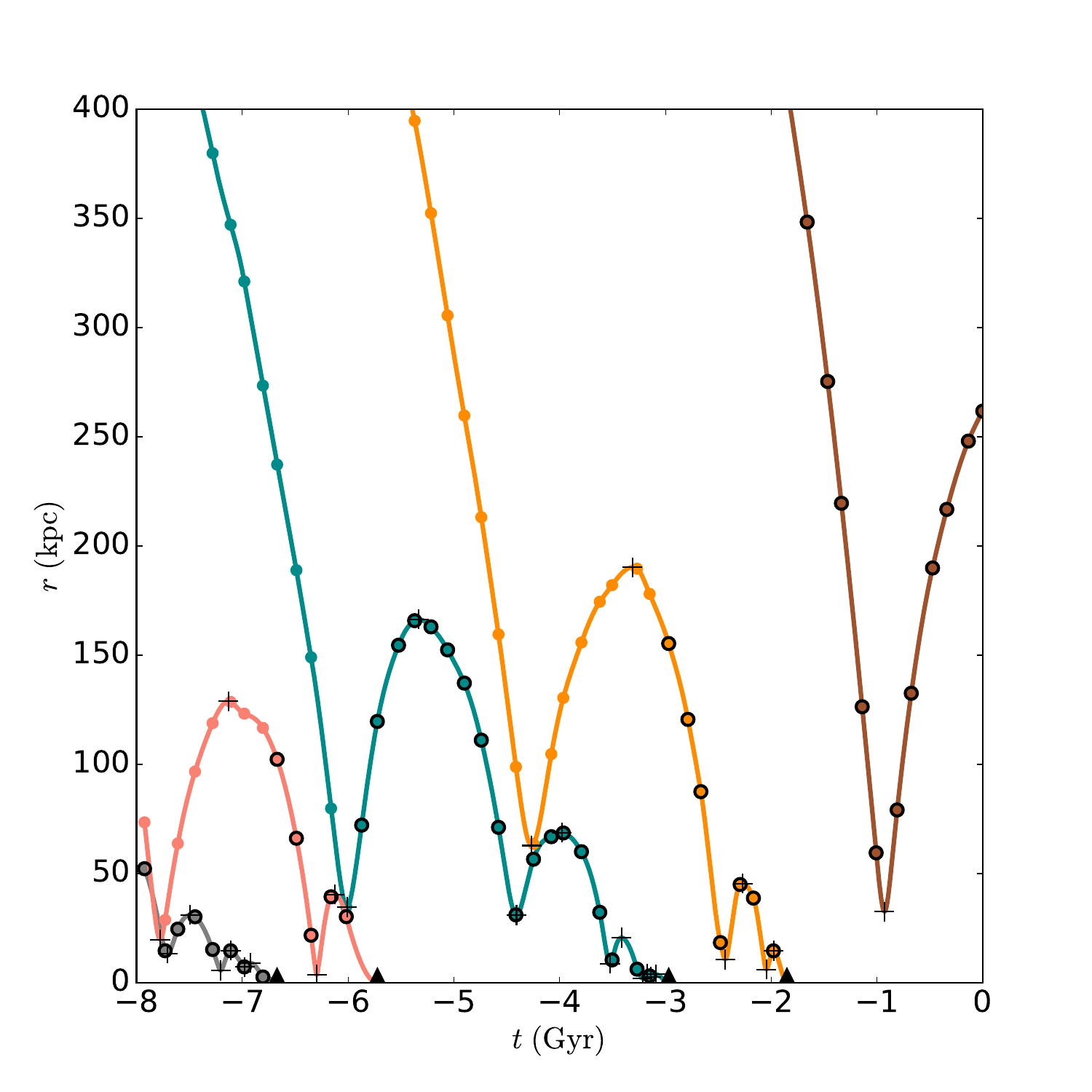}}}}
\caption{An example of a galaxy that has five different closest companions between z=1 and z=0.
  The orbits of all five companions are shown, using a different colour for each orbit.
  Within each orbit, the snapshots are depicted with filled circles, while the 6D interpolation
  is shown with a solid line.  Pericentres and apocentres are identified using {\bf +} symbols.
  For the four orbits that end in a merger, the time of the merger is depicted with a black triangle.
  At every snapshot in time, the closest companion is highlighted using a black circle.
\label{figorbitcc}}
\end{figure}

Figure~\ref{figorbitcc} provides an illustration of the latter scenario, showcasing a galaxy that has five different
closest companions
over the redshift range of $0 \le z < 1$.  The reconstructed orbits of all of these companions are shown, with black circles
used to identify the closest companion at each snapshot.
Four of these companions merge with the given galaxy after undergoing at least
two close encounters.  After each of these mergers, the closest companion designation switches to a different companion
that was further out and which was already on a decaying orbit.  

\section{Close Encounters and Mergers}\label{secchar}

In this section, we move from the examination of individual reconstructed orbits to a statistical analysis of the
orbit sample as a whole.  Our focus in this section is to study the prevalence and nature of close encounters and mergers 
in the \citet{patton20} sample of closest-companion galaxy pairs in IllustrisTNG.

\subsection{Close Encounters}\label{secencounters}

In assessing the extent to which a pair of galaxies may have been altered by a previous interaction,
two key questions arise: how long has it been since the galaxies had their last pericentre encounter (if ever),
and how close together (in space) were the galaxies at that time?  If the encounter
was sufficiently close, it is reasonable to expect that the galaxies may have undergone significant
disruptions as a result.  And if the encounter was relatively recent, one might expect that
signs of any such disruptions might still be detectable.

\subsubsection{The Times and Separations of Previous Encounters}

\begin{figure}
\centerline{\rotatebox{0}{\resizebox{11.0cm}{!}
{\includegraphics{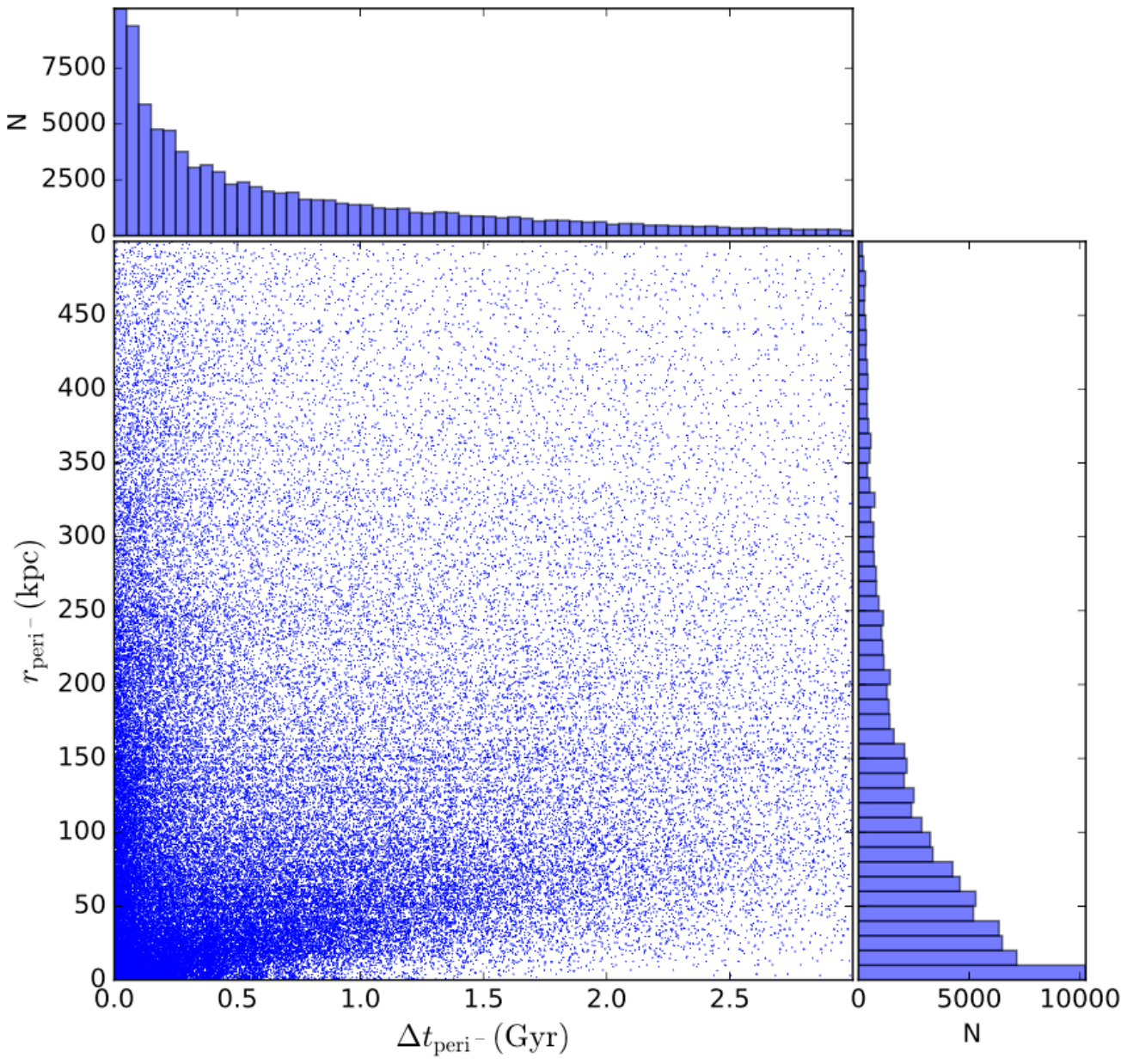}}}}
\caption{The 3D separation at the previous pericentre ($\rperip$) is plotted versus the time since
  the previous pericentre ($\dtperip$) in the main panel of this figure.  The histogram at the top shows the distribution of
  $\dtperip$ (summed over the given range in $\rperip$), while the histogram on the right
  shows the distribution of $\rperip$ (summed over the given range in $\dtperip$).  Closest companion pairs
  which have $\rperip > 500$ kpc or $\dtperip > 3$ Gyr  are relatively rare and are not shown.
\label{figperihist}}
\end{figure}

Having computed the time since the most
recent pericentre ($\dtperip$) and the pair separation at that pericentre ($\rperip$) for every pair
in our sample (see Section~\ref{secperiapo}), we can tackle both of the questions posed in the preceding
paragraph.
In Figure~\ref{figperihist}, we plot $\rperip$ versus $\dtperip$ for 
the 93,926 galaxies that have experienced a closest companion encounter within 500 kpc during the past 3 Gyr.
The majority of these encounters have occurred relatively recently (within the past Gyr) and at
separations that are sufficiently close (within about 100 kpc) that the galaxies in the pair would likely have had
substantially overlapping dark matter haloes during the encounter.  This suggests that
a substantial fraction of the closest companion pairs in our sample may have been altered by
relatively recent encounters.

\subsubsection{The Previous Pericentre Fraction ($f_{\rm peri}$)}

While the distributions of $\rperip$ and $\dtperip$ shown in Figure~\ref{figperihist} are instructive,
more information is needed if we are to quantify the prevalence of close encounters in our sample.
In particular, we must account for the fact that there is a limit to how far into the past we can track
the orbit of each galaxy pair, and this will limit our ability to detect encounters at any given time in the past.
In order to quantify the prevalence of recent close encounters in our sample of closest companion pairs,
we therefore set out to compute the fraction of galaxies that have had a previous pericentre encounter with their closest
companion (hereafter $f_{\rm peri}$) as a function of lookback time (hereafter $t_L$).
Given the redshift limits of our pairs sample ($0 \leq z < 1$), we can track some orbits nearly 8 Gyr into the past (those
at very low redshift), while we have no available orbital history for pairs at snapshot 50 ($z \sim 1$).
We address this issue by computing $f_{\rm peri}$ at a range of $t_L$,
including only those orbits that have not yet dropped out of the sample at any given lookback time.

\begin{figure}
\centerline{\rotatebox{0}{\resizebox{9.0cm}{!}
{\includegraphics{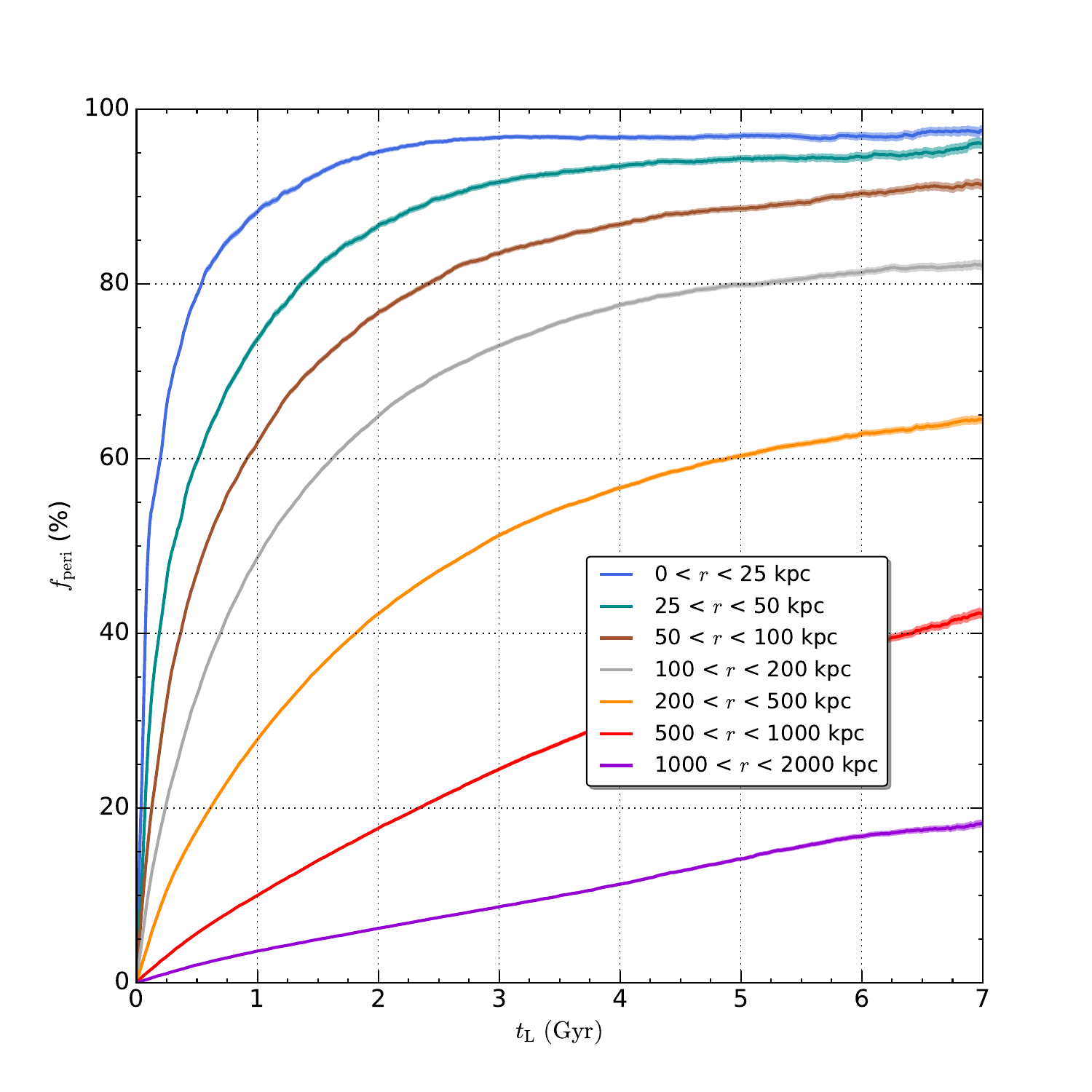}}}}
\caption{The fraction of galaxies that have had a previous encounter with their closest companion ($f_{\rm peri}$) is plotted
  versus lookback time ($t_L$) for different ranges of 3D closest companion separations.
  The shaded regions depict $1\sigma$ binomial errors on $f_{\rm peri}$.
\label{figfperi}}
\end{figure}

In Figure~\ref{figfperi}, we plot $f_{\rm peri}$ versus lookback time for different ranges of 3D closest companion separations ($r$).
Pairs with smaller separations have the highest average $f_{\rm peri}$ at all times.  This is to be expected, since pairs
with small $r$ are more likely to be found near the pericentre of their orbit, and pairs with small $r$ are also more likely
to be found on smaller orbits with shorter orbital timescales.
The pericentre fraction rises rapidly with $t_L$ for the closest pairs ($r < 25$ kpc), reaching 88.2 $\pm$ 0.4 per cent by a
lookback time of 1 Gyr.  In other words, nearly every close pair has had a recent pericentre encounter
with its closest companion.  It is also notable that $f_{\rm peri}$ reaches a plateau of about 97 per cent at $3 < t_L < 7$ Gyr.
This suggests that the remaining 3 per cent of close pairs never have a discernable first pericentre (e.g. if they merge
very soon after their first close approach on a nearly radial orbit).

At wider pair separations, $f_{\rm peri}$ drops quickly as a function of increasing $r$, at any given lookback time.
For example, only 49 per cent of pairs with $100 < r < 200$ kpc have had a pericentre encounter within the past Gyr,
and only 10 per cent of pairs with $500 < r < 1000$ kpc have had a pericentre encounter within the past Gyr.
And at the widest separations that we track (1000-2000 kpc), fewer than 20 per cent of the pairs have had a pericentre
encounter within the past 7 Gyr.  In summary, larger pair separations are associated with a decreased likelihood of the pair
having had a recent pericentre encounter.  

\subsubsection{Merging vs. Non-merging Pairs}

Our sample of reconstructed orbits contains a wealth of information about the types of interactions that
take place between galaxies in IllustrisTNG.  As seen in Figure~\ref{figdiversity}, some closest-companion
pairs are on merging orbits, while others appear unlikely to merge soon (if ever).  We can assess the
relative proportions of merging and non-merging pairs in the sample as a whole by comparing
the pericentre separations of the encounters that precede and follow each galaxy pair.  For a sample of
merging pairs, we would expect that the separation at the next encounter ($\rperin$) would
typically be smaller than the separation at the preceding encounter ($\rperip$).  Conversely,
for a sample of non-merging pairs, we would expect no significant difference between $\rperip$ and $\rperin$.

\begin{figure}
\centerline{\rotatebox{0}{\resizebox{11.0cm}{!}  
{\includegraphics{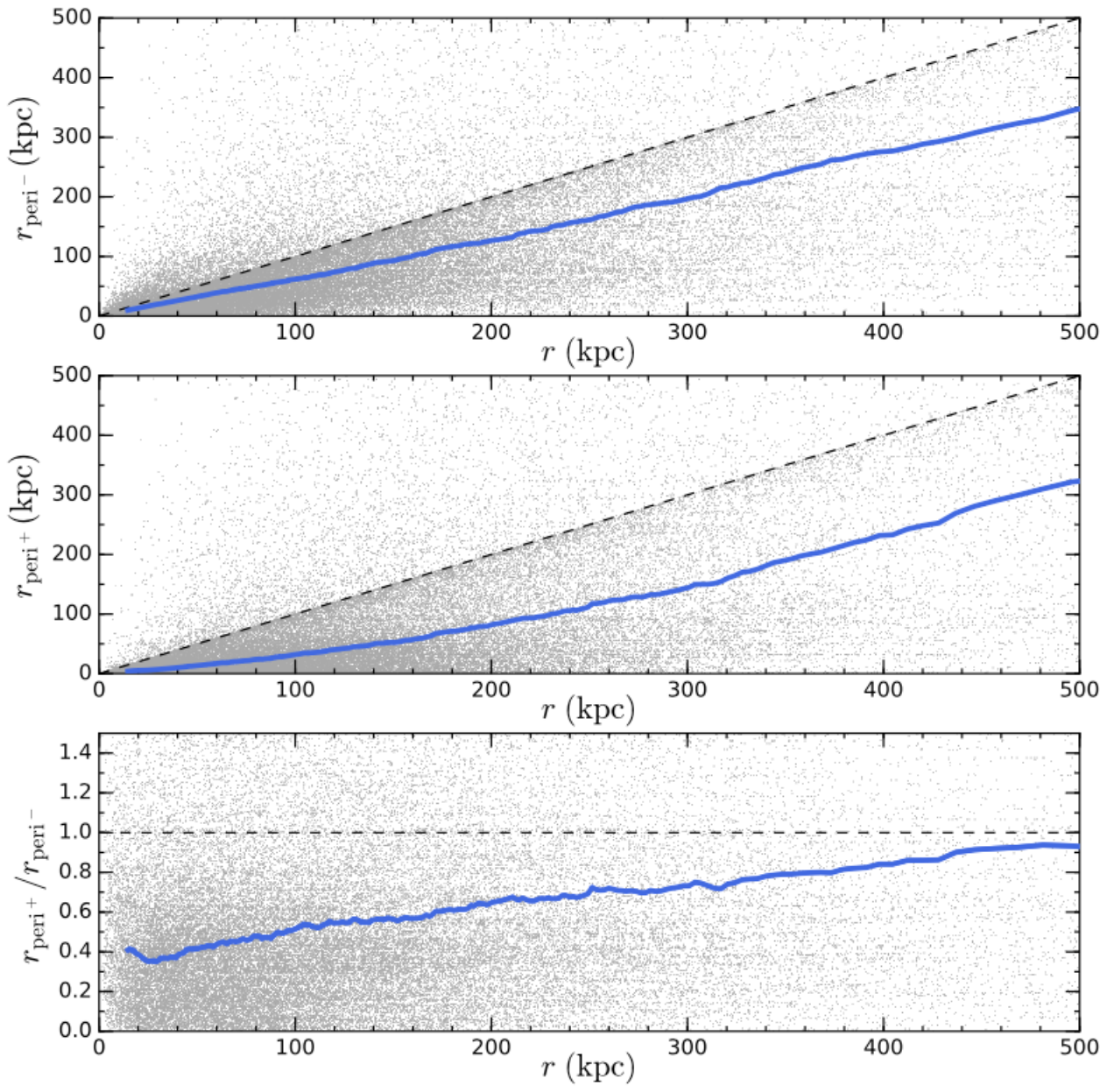}}}}
\caption{
  In the upper panel, the separation at the previous pericentre ($\rperip$) is plotted vs. the 3D
  distance to the closest companion ($r$).  The dashed line denotes $r = \rperip$, and the
  solid blue line represents the median $\rperip$ (with the line width corresponding to the 1$\sigma$ uncertainty
  in the median).
  In the middle panel, the separation at the next pericentre ($\rperin$) is plotted vs. $r$, with the dashed line
  denoting $r = \rperin$ and the solid blue line representing the median $\rperin$.
  In the lower panel, $\rperin/\rperip$ is plotted vs. $r$, with the dashed line denoting $\rperin = \rperip$
  (i.e. no change in minimum separation from one pericentre to the next), and the solid blue line representing
  the median $\rperin/\rperip$.
\label{figfrperipn}}
\end{figure}

In Figure~\ref{figfrperipn}, we explore the relationship between $\rperip$ and $\rperin$ by plotting each of these
quantities versus pair separation (in the upper and middle panels respectively) and by plotting $\rperin/\rperip$ versus
pair separation in the lower panel.  We restrict this sample to pairs with separations less than 500 kpc
that have at least one pericentre in their past {\it and} in their future, thereby including the same set of
galaxy pairs in all three panels.

In the upper panel of Figure~\ref{figfrperipn}, we can see that $\rperip$ increases as a function of $r$.
In addition, while $\rperip$ is usually smaller than $r$, there is a distinct subset of close pairs that have
$\rperip > r$, which is what we would expect in a shrinking and potentially merging orbit
(see e.g. ``Orbit a'' in Figure~\ref{figdiversity}).
In the middle panel, wider pairs are found to have (on average) larger $\rperin$.  In addition,
$\rperin$ is rarely larger than $r$, especially at small separations. In these cases,
the orbits are increasing in size (see e.g. ``Orbit f'' in Figure~\ref{figdiversity}).

In the lower panel of Figure~\ref{figfrperipn}, we plot $\rperin/\rperip$ versus pair separation, and compare
with the null hypothesis of $\rperin/\rperip \sim 1$ that would be expected in the absence of mergers.
At small separations (roughly $r < 50$ kpc), the median $\rperin/\rperip$ is approximately $0.4$; that is,
the median pericentre decreases by about 60 per cent from one encounter to the next.  This implies that many
of these close pairs are on rapidly shrinking merging orbits.  At larger separations, the median $\rperin/\rperip$ 
increases, nearly reaching the null hypothesis by $r \sim 500$ kpc.  This suggests that closest companion pairs
with separations larger than about 0.5 Mpc are dominated by galaxies on non-merging or slowly decaying orbits.  

\subsection{Mergers}\label{secmergers}

For the reconstructed orbits in which the pairs merge, we define the time of the merger ($\tmerge$) to be
the time of the first snapshot when the two galaxies have the same descendant (see \S~\ref{secassembly}).
By comparing this value with the time at which the pair is identified, we can compute the time until the pair merges.
For pairs at the high redshift end of our sample, we can track pairs for nearly 8 Gyr into the future, providing plenty of
opportunity to see if the pair merges.  For pairs near $z \sim 0$, there is instead only a short window of time
within which a merger may occur.

\begin{figure}
\centerline{\rotatebox{0}{\resizebox{9.0cm}{!}
{\includegraphics{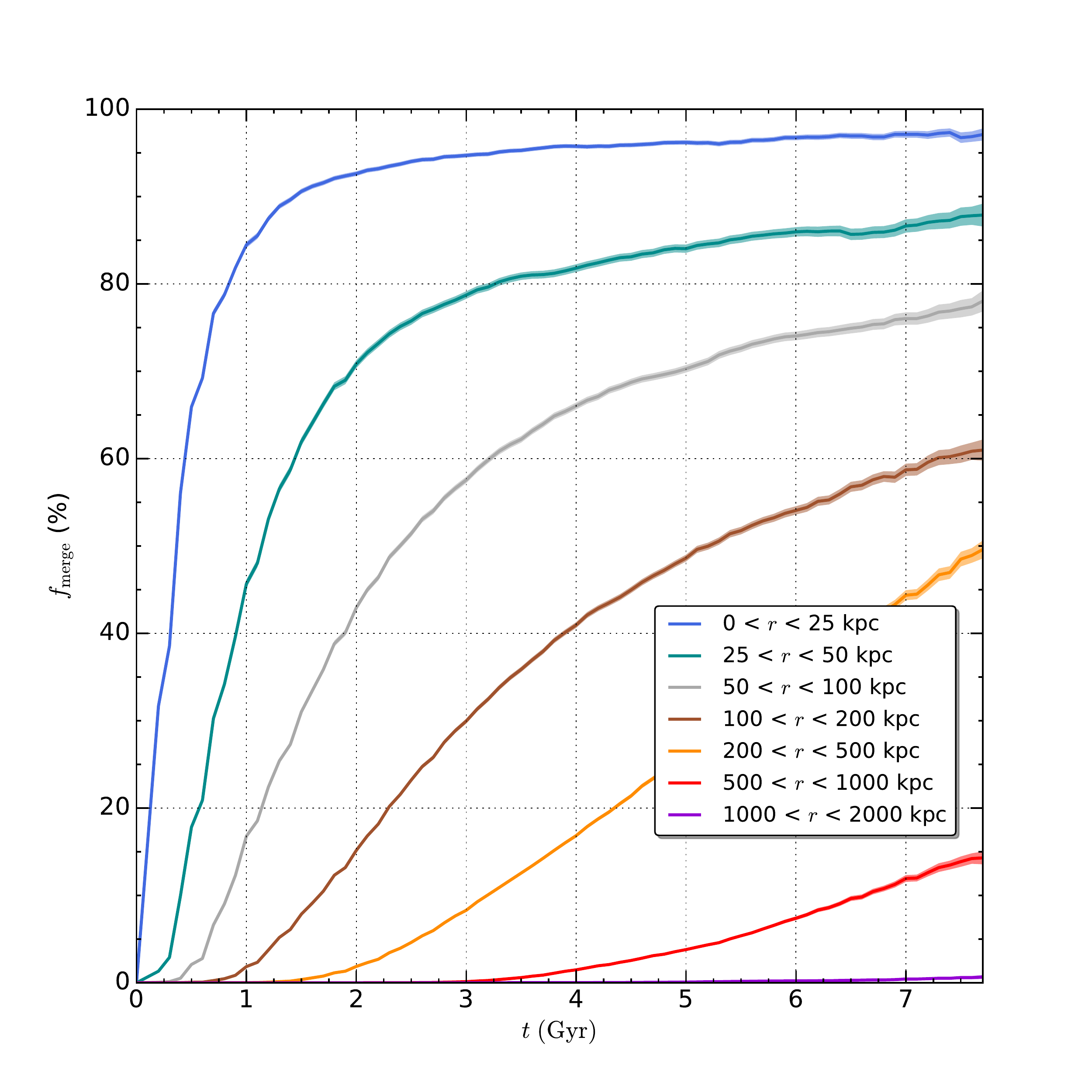}}}}
\caption{The merger fraction ($\fmerge$) is plotted versus time 
  for subsets of the sample divided according
  to the 3D distance to the closest companion, as specified in the legend.
  The shaded regions depict $1\sigma$ binomial errors on the merger fraction.
  The majority of galaxies with very close companions merge quickly.
  At larger separations, mergers are increasingly rare and occur much later.  
\label{figfmerge}}
\end{figure}

Following the same methodology that was used in Section~\ref{secencounters},
we compute the fraction of pairs that will merge (hereafter $\fmerge$) as a function of time, considering only those
systems that have not yet dropped out of the sample at any given time.
In Figure~\ref{figfmerge}, we plot $\fmerge$ versus time, for different subsets of 3D pair separation.
We find that $\fmerge$ rises rapidly for
the closest pairs ($r < 25$ kpc), reaching 84.5 $\pm$ 0.4 per cent within 1 Gyr, and 92.6 $\pm$ 0.3 per cent within 2 Gyr.
On longer timescales ($> 5$ Gyr), $\fmerge$ appears to plateau at about 97 per cent, suggesting that roughly 3 per cent
of the closest pairs are flyby systems that never merge.  At larger pair separations, the merger fraction becomes progressively
smaller at any given time.
In addition, there is an increasingly large time delay before $\fmerge$ begins to rise for the wider pairs, 
as its may take billions of years for the galaxies to reach each other in pairs with large separations.

\begin{figure}
\centerline{\rotatebox{0}{\resizebox{9.0cm}{!}
{\includegraphics{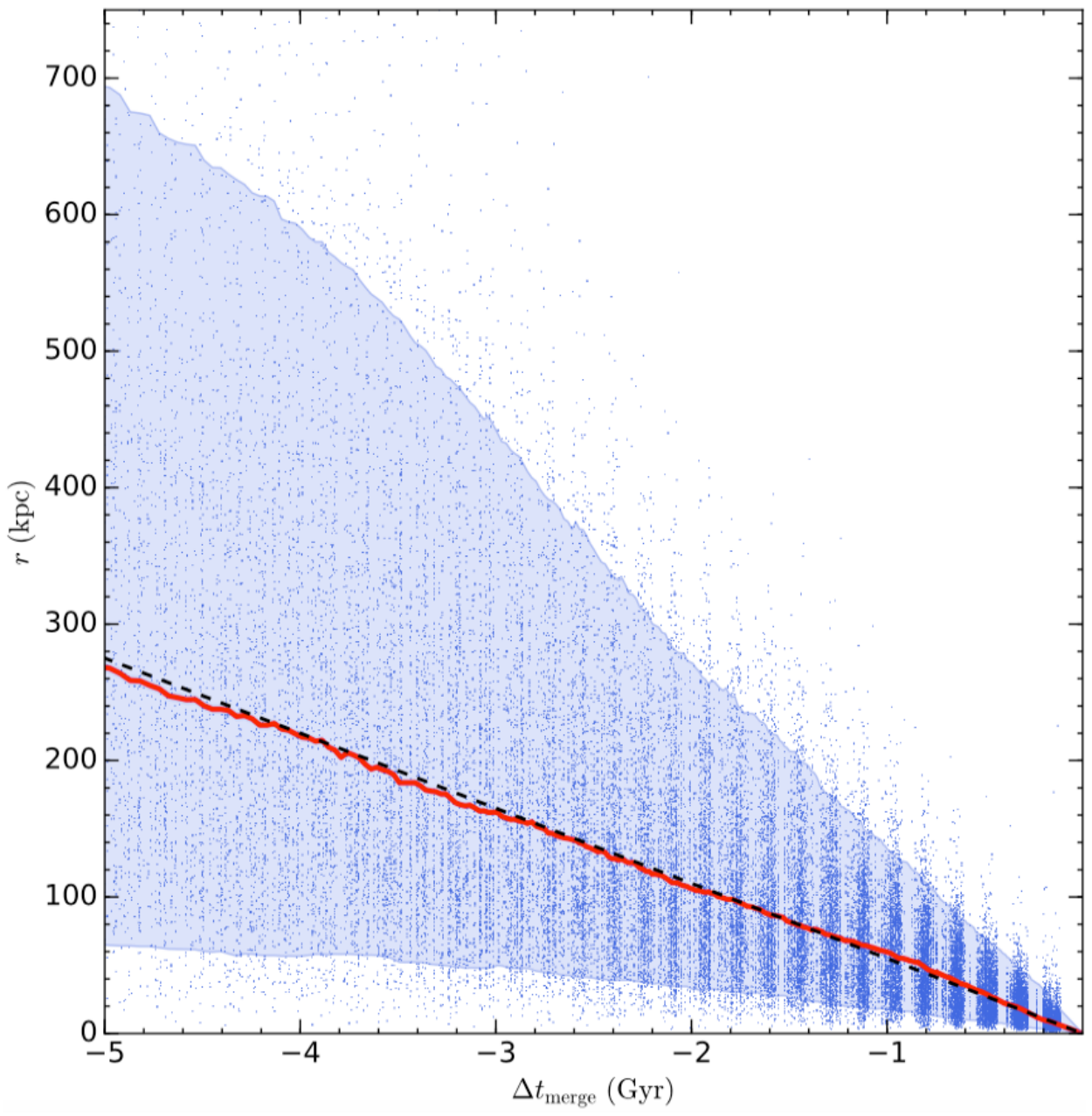}}}}
\caption{Pair separation ($r$) is plotted versus time relative to the merger ($\dtmerge$) for closest companion
  pairs that merge by $z=0$.  The thick red line denotes median $r$, and the thickness of the line
  depicts the $1\sigma$ uncertainty on the median.
  The blue shaded region encloses the 95th and 5th percentile in $r$.
  The black dashed line corresponds to a linear fit to the median, with a slope of $-$55 kpc/Gyr.
\label{figrmerge}}
\end{figure}

In Figure~\ref{figrmerge}, we examine the relationship between pair separation and time relative to the merger
(hereafter $\dtmerge$).
Pairs that will merge in 1 Gyr have a median separation of 60 kpc, and 90 per cent of these pairs have separations of
18-136 kpc.
On longer timescales, pairs that will merge in 5 Gyr have a median separation of 268 kpc, and 90 per cent of these pairs
have separations of 65-694 kpc.
A linear fit to the median separation versus $\dtmerge$ has a slope of $-$55 kpc/Gyr, giving us a sense of the typical
time until the merger at a given separation (for merging pairs).  

\section{Conclusions and Future Work}\label{secconclusions}

We have reconstructed the orbits of massive galaxies ($M_* > 10^{10} \msun$) 
and their closest companions in the highest resolution run
of the 100 Mpc IllustrisTNG cosmological simulation (TNG100-1).
For each closest companion pair at $z < 1$ 
identified by \citet{patton20},
we trace the orbit backwards in time (to a redshift of $z \sim 1$) and forwards in time (to $z = 0$),
using the 50 snapshots that are available.  
We interpolate between each pair of snapshots (which are 162 Myr apart on average) using a novel
kinematic interpolation technique, and we use the merger simulations of \citet{patton13} to show that
these 6D interpolations are much more effective at identifying and characterizing close encounters than
3D interpolations (which use separations in each of the $x$, $y$ and $z$ directions) or 
1D interpolations (which use only radial separations).

After analysing the resulting sample of reconstructed orbits, our main conclusions are as follows:
\begin{enumerate}
\item IllustrisTNG closest companion pairs include a diverse mix of rapidly merging pairs, slowly merging pairs, long lived satellites,
flyby encounters, pairs in groups and clusters, etc.
\item almost 90 per cent of the closest pairs ($r < 25$ kpc) have undergone a pericentre encounter within the past Gyr
\item galaxies in the closest pairs are often found on rapidly shrinking orbits, and roughly 85 per cent of these pairs will merge within 1 Gyr
\item roughly 3 per cent of the closest pairs appear to be flyby systems that will never merge
\item the median separation of merging pairs decreases by $\sim$ 55 kpc/Gyr during the final 5 Gyr preceding the merger
\end{enumerate}

The results and trends reported here are likely to depend on various properties of the galaxy pairs, including the stellar mass and mass ratio of the galaxies, the environments that the pairs reside in, and the redshifts of the pairs.  While our sample is large enough to investigate these dependencies, this analysis is beyond the scope of this study and is left to a future paper.

The reconstructed orbits presented in this work can also be used to study how galaxy properties change during
close encounters, enabling us to uncover the physical processes that give rise to the correlations
between galaxy properties and pair separation that have been reported in observed samples of galaxy pairs.
By following merging galaxy pairs from their earliest close encounters through to their mergers and beyond, it will be possible to
draw more direct connections between studies of pre-merger and post-merger galaxies in cosmological simulations.
In addition, the properties of the orbits themselves (such as eccentricity and orbital period) can be used
to analyze the dynamics of interacting galaxies in a realistic cosmological context, potentially relating
the characteristics of these pre-merger orbits to the properties of the post-merger galaxies.

Finally, the kinematic (6D) interpolation scheme introduced in this study can also be applied to other cosmological simulations,
allowing researchers to reconstruct the orbits of galaxies in between the available snapshots.  While we have
designed this approach with interacting galaxies in mind, these methods should be equally useful for researchers
wishing to reconstruct the orbits of galaxies in groups and clusters of galaxies.

\section*{Acknowledgements}

We thank the anonymous referee for an insightful and constructive report.  
We thank all members of the IllustrisTNG collaboration for making their data publicly available.
DRP and SLE gratefully acknowledge NSERC of Canada for Discovery Grants which helped to fund this research.
RIW was supported in part by an NSERC Undergraduate Student Research Award.
We thank Westley Brown and Shoshannah Byrne-Mamahit  for helpful discussions.  

\section*{Data Availability}

The IllustrisTNG data used in this work are publicly available at https://www.tng-project.org.





\bsp	
\label{lastpage}
\end{document}